\documentclass[apj]{emulateapj}
\usepackage{amsmath}
\usepackage{epstopdf}
\usepackage{bm}
\usepackage{natbib}

\usepackage{color}

\shorttitle{Bar and Spiral Patterns of NGC 1365}
\shortauthors{Speights \& Rooke}

\begin{document}

\title{The Dynamical Relationship Between the Bar and Spiral Patterns of NGC 1365}

\author{Jason C. Speights \& Paul C. Rooke}
\affil{Department of Physics and Engineering, Frostburg State University, Frostburg, MD 21532, USA}
\email{jcspeights@frostburg.edu}
  
\begin{abstract}

\small Theories that attempt to explain the dynamical relationship between bar and spiral patterns in galactic disks make different predictions about the radial profile of the pattern speed.  These are tested for the H-alpha bar and spiral patterns of NGC 1365. The radial profile of the pattern speed is measured by fitting mathematical models that are based on the Tremaine-Weinberg method. The results show convincing evidence for the bar rotating at a faster rate than the spiral pattern, inconsistent with a global wave mode or a manifold. There is evidence for mode coupling of the bar and spiral patterns at the overlap of corotation and inner Lindblad resonances, but the evidence is unreliable and inconsistent.  The results are the most consistent with the bar and spiral patterns being dynamically distinct features. The pattern speed of the bar begins near an ILR and ends near the corotation resonance. The radial profile of the pattern speed beyond the bar most closely resembles what is expected for coupled spiral modes and tidal interactions.  

\end{abstract}

\keywords{\small galaxies: evolution -- galaxies: individual (NGC 1365) -- galaxies: kinematics and dynamics  -- galaxies: spiral -- galaxies: structure -- methods: data analysis}

\section{INTRODUCTION}

\small The evidence for a dynamical relationship between the bar and spiral patterns of galactic disks is widely recognized.  Well organized spiral patterns are common in barred galaxies (Kormendy \& Norman 1979).  Bars can provide periodic forcing of spiral patterns (Sanders \& Huntley 1976, Huntley et al. 1978, Schwarz, M. P. 1981).  The presence of a strong bar approximately doubles the likelihood of a field galaxy possessing a grand design spiral pattern (Elmegreen \& Elmegreen 1982).  Maximum bar and spiral pattern strengths are correlated for local galaxies (Salo et al. 2010). 

Despite the evidence for its existence, the nature of the relationship remains unclear (e.g., Sellwood \& Sparke 1988, Buta et al. 2009).  Theories that attempt to explain it make different predictions about the radial profile of the pattern speed, $\Omega_p$.  In this context they are classifiable according to three general categories (Dobbs \& Baba 2014).   \\[4pt]
\indent \begin{minipage}[ht!]{0.01\textwidth} \vskip -0pt 
(1)  
\end{minipage} 
\hskip 2pt \begin{minipage}[h!]{0.43\textwidth} \vskip 0pt
The bar and spiral patterns have the same value of $\Omega_p$. 
\end{minipage}\\
\indent \begin{minipage}[ht!]{0.01\textwidth} \vskip -7.5pt 
(2)  
\end{minipage} 
\hskip 2pt \begin{minipage}[h!]{0.43\textwidth} \vskip 2pt
The bar and spiral patterns have different values of $\Omega_p$, and are coupled by resonance.
\end{minipage}\\
\indent \begin{minipage}[ht!]{0.01\textwidth} \vskip -7.5pt 
(3)  
\end{minipage} 
\hskip 2pt \begin{minipage}[h!]{0.43\textwidth} \vskip 2pt
The bar and spiral patterns have different values of $\Omega_p$, and are dynamically distinct features. 
\end{minipage}\\[1pt]

The purpose of this paper is to test these predictions for the H-alpha (H$\alpha$) bar and spiral patterns of NGC 1365.  The bar and spiral patterns of NGC 1365 are among the most studied galactic structures (e.g., Lindblad 1999).  Most authors assume category (1) (e.g., J\"{o}rs\"{a}ter \& van Moorsel 1995, hereafter JM95; Lindblad et al. 1996, hereafter L96; Canzian 1998; Vera-Villamizar et al. 2001, hereafter V01; Pi\~nol-Ferrer et al. 2012).  Although photometric features of NGC 1365 are often cited as evidence for category (1), the gas kinematics are inconsistent with this category (Sellwood \& Wilkinson 1993, and references therein).  Furthermore, measurements of $\Omega_p$ for the H${\scriptsize \mbox{I}}$ spiral pattern are consistent with an $\Omega_p$ that decreases with increasing radius (Speights \& Westpfahl 2011, hereafter SW11).

All three categories include density wave theories (e.g., Lindblad 1963; Lin \& Shu 1964, 1966; Rohlfs 1977 and Bertin \& Lin 1996 for reviews).  They assume $\Omega_p$ is constant with increasing radius (i.e., a rigid pattern), in which case there may exist corotation resonance (CR), inner Lindblad resonance (ILR), and outer Lindblad resonance (OLR).  The CR occurs when $\Omega_p$ = $\Omega$, the angular frequency of an orbit.  The ILR and OLR occur when $\Omega_p$ = $\Omega$ $\pm$ $\kappa$/$m$, where $\kappa$ is the epicycle frequency of an orbit, and $m$ is an integer determined by the symmetry of the pattern.    

Category (1) includes global wave modes with the properties described above, and manifolds (Romero-G\'{o}mez et al. 2006, Athanassoula 2012 for a review).  Simulations of bar driven spiral patterns sometimes resemble global wave modes (e.g., Sanders \& Huntley 1976, Huntley et al. 1978, Schwarz, M. P. 1981, Roca-F{\`a}brega et al. 2013).  Manifolds guide material away from unstable Lagrange points near the ends of the bar, along paths resembling spiral patterns.   

Category (2) assumes that the radius of the CR for a faster bar is approximately coincident with the radius of the ILR for a slower spiral density wave (Tagger et al. 1987, Sygnet et al. 1988).  The ILR for $m$ = 2 and 4 is especially important for mode coupling.  The coupling of these resonances can drive a spiral density wave.  
  
Category (3) allows for many possible explanations for spiral patterns.  The most common are: kinematic density waves (Lindblad 1956, Kalnajs 1973); material arms (e.g., Wada et al 2011, Grand et al. 2012, Kawata et al. 2014); shearing instabilities that are swing amplified (Goldreich \& Lynden-Bell 1965; Julian \& Toomre 1966); spiral modes that are overlapping (Sellwood \& Carlberg 2014), or coupled as described for category (2); and tidal interactions with other galaxies (Toomre 1969, Kormendy \& Norman 1979), or halo substructures and dark satellites (Tutukov \& Fedorova 2006, Dubinski et al. 2008).  

The most common explanations for spiral patterns in category (3) are distinguishable by the expected results for the radial profile of $\Omega_p$.  For kinematic density waves, the profile is expected to approximately follow the possible locations for ILR.  For material arms and shearing instabilities that are swing amplified, the profile is expected to approximately follow $\Omega$.  For overlapping modes, the profile is also expected to approximately follow $\Omega$ because the methods used in this paper average the overlapping values of $\Omega_p$.  For coupled spiral modes, the profile is expected to resemble a step function, and occur between the locations for ILR and CR.  For tidal interactions, the profile is expected to differ from $\Omega$, but show a shear rate that is approximately similar to that of $\Omega$  (e.g., Oh et al. 2008, Dobbs et al. 2010).

A full description of the relevant theories is beyond the scope of this paper.  For a review of bar patterns see Sellwood \& Wilkinson (1993).  For a review of spiral patterns see Dobbs \& Baba (2014).  

The radial profile of $\Omega_p$ is measured by fitting mathematical models that are based on the Tremaine \& Weinberg (1984, hereafter TW84) method.  Two additional models are fit to assist with analyzing the results.  A model of the kinematics is fit to H$\alpha$ velocity field data for calculating $\Omega$ and other possible locations for resonance.  A model of the approximate location of the gravitational potential minimum is fit to K$_{\mbox{\footnotesize s}}$ band data for distinguishing the locations of the bar and spiral patterns.

The rest of this paper is organized as follows.  Section 2 describes the data.  Section 3 defines the mathematical models.  Section 4 explains the model fitting methods.  Section 5 presents the results.  Section 6 discusses the results.  Section 7 is a summary.

\section{DATA}

The H$\alpha$ data are maps of intensity, $I$, and line-of-sight velocity deprojected onto the galaxy disk, $V_{\footnotesize y}$.  The values of $V_{\footnotesize y}$ are calculated from the line-of-sight velocity, $V_{\mbox{\footnotesize los}}$, the systemic velocity, $V_{\mbox{\footnotesize sys}}$, and the disk inclination angle, $\phi_i$,
\begin{equation}
V_{\footnotesize y} = \frac{V_{\mbox{\footnotesize los}} - V_{\mbox{\footnotesize sys}}}{\mbox{sin}(\phi_{\footnotesize i})}.
\end{equation}
The uncertainties for $V_{\footnotesize y}$ are calculated by assuming the values for $V_{\mbox{\footnotesize sys}}$ and $\phi_i$ are well known and uncorrelated, and propagating the remaining uncertainties for $V_{\mbox{\footnotesize los}}$ through Equation (1),
\begin{equation}
\sigma_{V\footnotesize y} = \frac{\sigma_{V\mbox{\footnotesize los}}}{\mbox{sin}(\phi_{\footnotesize i})}.
\end{equation}
The values of $V_{\mbox{\footnotesize sys}}$ and $\phi_i$ are discussed in Section 3.1.

The data for $I$, $V_{\mbox{\footnotesize los}}$, and their associated uncertainties, are from Z\'anmar S\'anchez et al. (2008, hereafter Z08).  They observed NGC 1365 using the Rutgers Imaging Fabry-Perot interferometer at the Cerro Tololo Inter-American Observatory (CTIO)\footnote{CTIO is a division of the National Optical Astronomy Observatories, which are operated by the Association of Universities for Research in Astronomy, Inc., under contract with the National Science Foundation.}, and then fit Voigt profiles to the data cube. A Voigt profile is a combination of a Lorentz profile describing the natural line shape, and a Gaussian profile accounting for the effects of thermal broadening (Kwok 2007, Chapter 5). Z08 improve the signal-to-noise ratio before fitting the Voigt profiles by combining adjacent pixels in bins of 3 $\times$ 3 pixels for intermediate strength emission, and bins of 5 $\times$ 5 pixels for weak emission.  The pixel size is 0\farcs 98.  The approximate seeing is 2\farcs 4.  More details about the observations and data reduction are provided by Z08.

The data are shown in Figure 1.  Included in the figure are data in the optical B and near infrared K$_{\mbox{\footnotesize s}}$ bands for comparison.  The details about the observations and data reduction for the B band data are provided by Kuchinski et al. (2000), and for the K$_{\mbox{\footnotesize s}}$ band image by Jarrett et al. (2003).  Sources of starlight from the Milky Way are identified for removal from the K$_{\mbox{\footnotesize s}}$ band data using Source-Extractor  (Bertin \& Arnouts 1996).  

\begin{figure*}
\begin{centering}
\includegraphics[width=1\textwidth]{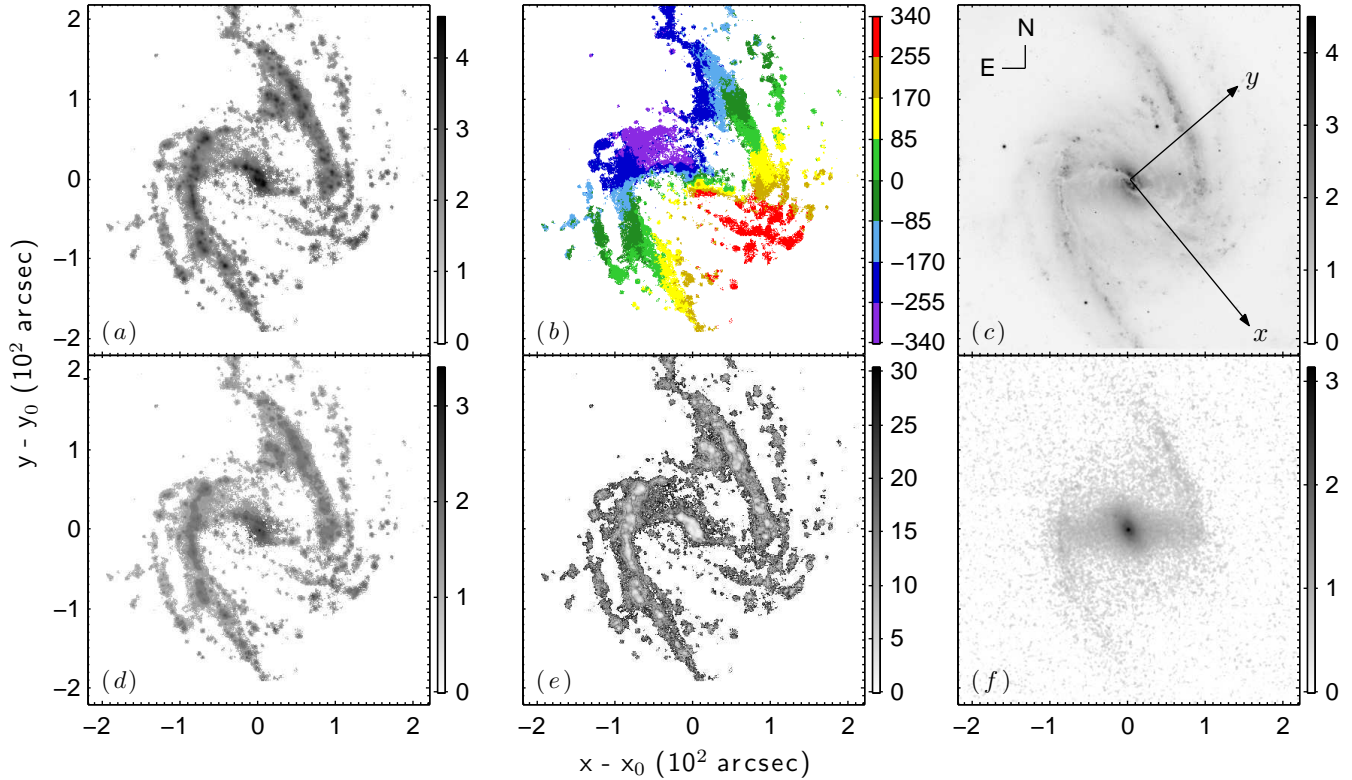} 
\caption{Data for NGC 1365.  The panels show: ($a$) $I$, ($b$) $V_{\footnotesize y}$, ($c$)  B band, ($d$) $\sigma_I$, ($e$) $\sigma_{\mbox{\footnotesize V}y}$, and ($f$) K$_{\mbox{s}}$ band.  In panels ($a$), ($c$), ($d$), and ($f$), the data are log scaled, and the colorbars are in units of log(counts). In panels ($b$) and ($e$), the colorbars are in units of km s$^{-1}$. In panel ($a$), the peak intensity is 4.0 $\times$ 10$^4$ counts.  In panel ($b$) the velocities are binned in increments of 85 km s$^{-1}$.  In panel ($d$), and the peak uncertainty is 2.6 $\times$ 10$^3$ counts. In panel ($e$) the peak uncertainty is 30.5 km s$^{-1}$. The receding half of the galaxy is to the bottom right. The coordinates x - x$_0$ and y - y$_0$ along the horizontal and vertical axes, respectively, are distances from the kinematic center. Panel ($c$) includes the sky coordinates east (E) and north (N), as well as the directions of the galaxy coordinates, $x$ and $y$, that are along the kinematic major and minor axis, respectively. The arrows pointing in the direction of positive $x$ and $y$ extend to the beginning of the warp at 240$^{\prime\prime}$.}
\end{centering}
\end{figure*}

\section{MATHEMATICAL MODELS}

\subsection{Adopted Disk Parameters}

\begin{deluxetable}{ll}[t!]
\tablecaption{Adopted Disk Parameters}
\tablewidth{0pt}
\startdata
\tableline \tableline
Parameter & Value \\
\tableline
 $\phi_i$  &  41$^\circ$  \\
 $\phi_p$  &  220$^\circ$  \\
 $V_{\mbox{\footnotesize sys}}$  & 1632 km s$^{-1}$   \\
 Kinematic center R.A. (J2000) & 03$^h$ 33$^m$ 36 \hskip -1.5 pt \fs 4 \\
  Kinematic center decl. (J2000) & -36$^\circ$ 08$^\prime$ 4$^{\prime\prime}$  
\enddata 
\end{deluxetable}

The models assume an approximately flat disk. The disk inclination, $\phi_i$, position angle, $\phi_p$,  $V_{\mbox{\footnotesize sys}}$, and location of the kinematic center, are adopted from Z08. The values of the adopted disk parameters are summarized in Table 1.  They are consistent with most other measurements (Sandqvist et al. 1982, JM95, L96), and are therefore assumed to be well known.  There is a warp in the disk that starts at $r$ $\sim$240$^{\prime\prime}$ (JM95), but only a small amount of the data are outside this radius.  The effect of these data are noted in Section 5.

\subsection{The Velocity Field Model}

A model of the velocity field is needed for measuring the azimuthal (tangential) velocity, $V_{\footnotesize \theta}$, which is used for calculating $\Omega$ = $V_{\footnotesize \theta}$/$r$, and the other possible locations for resonance.  The equation for the model is, 
\begin{equation}
\begin{split}
V_{\footnotesize y}(r,\theta) = V_{\footnotesize \theta}(r)\,\mbox{cos}(\theta).
\end{split}
\end{equation}
The angle $\theta$ is measured in the plane of the galaxy's disk, from the kinematic major ($x$) axis toward the minor ($y$) axis such that,
\begin{equation}
\mbox{cos}(\theta) = \frac{-(\mbox{x $-$ x$_{\mbox{\footnotesize 0}}$})\,\mbox{sin}(\phi_{\footnotesize p}) + (\mbox{y $-$ y$_{\mbox{\footnotesize 0}}$})\,\mbox{cos}(\phi_{\footnotesize p})}{r}.
\end{equation}
In Equation (4), the x and y coordinates are measured parallel to north and east, respectively.  The intersection of x = x$_0$ and y = y$_0$ is the location of the kinematic center (see Figure 1).  Fits of Equation (3) are found for concentric rings of data.\\

\subsection{The $\mathit \Omega_p$ Models}

The $\Omega_p$ models are based on the TW84 method.  The original form of the method has three assumptions: \\[4pt]
\indent \begin{minipage}[ht!]{0.01\textwidth} \vskip -1pt 
(1)  
\end{minipage} 
\hskip 4pt \begin{minipage}[h!]{0.9\textwidth} \vskip -3pt
The disk is flat. 
\end{minipage}\\
\indent \begin{minipage}[ht!]{0.01\textwidth} \vskip 0pt 
(2)  
\end{minipage} 
\hskip 4pt \begin{minipage}[h!]{0.9\textwidth} \vskip 0pt
The value of $\Omega_p$ is constant with radius.
\end{minipage}\\
\indent \begin{minipage}[ht!]{0.01\textwidth} \vskip -10pt 
(3)  
\end{minipage} 
\hskip 4pt \begin{minipage}[h!]{0.94\textwidth} \vskip 0pt
A tracer of the pattern obeys mass conservation in the \\ continuity equation. 
\end{minipage}\\[4pt]
The method integrates the continuity equation over an area of the disk bounded by $-\infty$ $<$ $x$ $<$ $\infty$ and $y_o$ $\leqslant$ $y$ $<$ $\infty$ (or similarly for $-y$). The result derived by TW84,
\begin{equation}
\int_{-\infty}^{+\infty}I(x,y_o)\,V_y(x,y_o)\,dx = \Omega_p\int_{-\infty}^{+\infty}I(x,y_o)\,x\,dx,
\end{equation}
relates $\Omega_p$ to the observables $I$ and $V_y$ for a tracer of the pattern. For ease of notation let, 
\begin{equation}
\mathcal{V}_y = \int_{-\infty}^{\infty}I(x,y_o)\,V_y(x,y_o)\,dx,
\end{equation}
and,
\begin{equation}
\mathcal{X} = \int_{-\infty}^{+\infty}I(x,y_o)x\,dx,
\end{equation}
so that Equation (5) is,
\begin{equation}
\mathcal{V}_y = \Omega_p\mathcal{X}.
\end{equation}
The high quality data used in this paper provide many calculations of $\mathcal{V}_y$ and $\mathcal{X}$ at different $y_o$ from which to fit a value for $\Omega_p$.  When these calculations are plotted, $\Omega_p$ is the slope of a line fit to $\mathcal{V}_y(\mathcal{X})$ that passes through $\mathcal{V}_y(0)$ = 0.  Equation (8) is also referred to in this paper as the original TW84 method.

For an $\Omega_p$ that varies with radius, Engstr\"oem (1994, hereafter E94) derives a general TW84 method,
\begin{equation}
\mathcal{V}_y = \int_{-\infty}^{+\infty}\Omega_p(r)\,I(x,y_o)\,x\,dx,
\end{equation}
from the phase-space continuity equation.  E94 goes on to show that due to the symmetry of the integrand this is equivalent to a Volterra equation of the 1st kind,
\begin{eqnarray} \nonumber
\hskip -22pt \mathcal{V}_y = \int_{y_o}^{+\infty}\Omega_p(r)&\hskip -72pt \big\{I(\sqrt{r^2-y_o^2},y_o) \\&\hskip 36 pt - I(-\sqrt{r^2-y_o^2},y_o)\big\}\,r\,dr,
\end{eqnarray}
by making the substitution $x$ = $\sqrt{r^2-y_o^2}$ in the right-hand side of Equation (9).\footnote{This result is rediscovered by Merrifield et al. (2006), but using a different derivation.} Equation (10) is also referred to in this paper as the general TW84 method.

To apply Equation (10) to the data, $\Omega_p$ is approximated as a piecewise constant function with $p$ values of $\Omega_p$ for $p$ concentric rings.  For a single path along $y_o$, the inner radius of the innermost ring is $y_o$.  In order to achieve convergence, the outermost ring has an inner radius that is set to 220$^{\prime\prime}$, and an outer radius that extends to the edge of the data.  

Fits of Equation (10) are unstable to noise unless the ring width is at least a few times larger than the angular resolution of the data (E94, Merrifield et al. 2006).  Stable fits for smaller ring widths are possible using regularization (Meidt et al. 2008a), but it is important to note that it is inappropriate to determine the functional form of $\Omega_p$ using only regularized fits because of the potential for a large amount of bias (Aster et al. 2005, Chapter 4).  Regularized fits can provide insight, though, about general trends in the radial behavior of $\Omega_p$ (Meidt et al. 2008b, 2009).  Both larger ring widths and regularization are used in this paper for obtaining stable fits. 

There are many examples of the TW84 method successfully applied to different components of the interstellar medium despite the likelihood of violating assumption (3).  Westpfahl (1998) applies the method to H${\scriptsize \mbox{I}}$ data by showing that the effect of the source function, $s(x,y)$, in the general form of the continuity equation is approximately negligible for a measurement of the instantaneous pattern speed.  Rand \& Wallin (2004) apply the method to CO data by arguing that the conversion between the different phases of the interstellar medium is negligible if the tracer used is the dominant component in a dynamical timescale.  Hernandez et al. (2005) apply the method to H$\alpha$ data by noting that sources of H$\alpha$ emission approximately trace the molecular gas density, and that H$\alpha$ velocity data traces the gravitational potential.  For H$\alpha$ data in particular, the TW84 method is also successfully applied by Emsellem et al. (2006), Fathi et al. (2007), Chemin \& Hernandez (2009), Fathi et al. (2009), and Gabbasov et al. (2009).

When assumption (3) is violated, it follows from the general form of the continuity equation that Equation (8) takes the form,
\begin{equation}
\mathcal{V}_y = \Omega_p^{\prime}\mathcal{X} + \mathcal{S},
\end{equation}
where that,
\begin{equation}
\mathcal{S} = \int_{y_o}^{+\infty}\int_{-\infty}^{+\infty}s(x,y)\,dx\,dy,
\end{equation}
is the integrated source function.  Although the functional form of $s(x,y)$ is unknown, Equation (11) shows that if violations of assumption (3) effect the results, the effect is detectable in the form of deviations from linearity in plots of $\mathcal{V}_y(\mathcal{X})$, or nonzero intercepts for the fitted lines.  

The effect of violating assumption (3) is investigated by comparing the results for Equations (8) and (11). The results presented in this paper show that calculated values of $\mathcal{V}_y(\mathcal{X})$ are well described by straight lines.  The value of  $\mathcal{S}$ is therefore approximated by assuming it is constant (i.e., $\mathcal{V}_y(0)$ = $\mathcal{S}$) in the fit of Equation (11).  The effect of violating assumption (3) is then quantified using the percent difference,
\begin{equation}
\bigtriangleup \Omega_p = \frac{\Omega_p - \Omega_p^\prime}{\Omega_p^\prime}.
\end{equation}
A similar analysis is excluded from the general TW84 method because unlike $\Omega_p$, for which it is reasonable to assume any azimuthal dependence is negligible, the azimuthal dependence of $s(x,y)$ is unknown.

\subsection{The Potential Minimum Model}

In addition to velocity field and $\Omega_p$ models, a model is fit to the K$_{\mbox{\footnotesize s}}$ band data for estimating the approximate locations of the gravitational potential minima associated with the bar and spiral patterns.  The results are used for comparison with the locations of changes in the radial behavior of $\Omega_p$.  

The model assumes $m$ = 2 rotational symmetry with respect to the kinematic center.  It has the form,
\begin{equation}
I(r,\theta)= I_{\footnotesize 0}(r) + I_{\footnotesize 2}(r)\,\mbox{cos}(2[\theta-\phi_{I2}(r)]), 
\end{equation}
for unknown parameters $I_{\footnotesize 0}(r)$, $I_{\footnotesize 2}(r)$, and $\phi_{I2}(r)$.  Although it is unlikely that variations in the intensity of the K$_{\mbox{\footnotesize s}}$ band data is exactly described by a cosine function, Equation (14) is adequate for its purpose.  

Equation (14) is nonlinear in the unknown parameters.  It is transformed into a linear equation,
\begin{equation}
I(r,\theta)= I_{\footnotesize 0}(r) + I_{\footnotesize 2x}(r)\,\mbox{cos}(2\theta) + I_{\footnotesize 2y}(r)\,\mbox{sin}(2\theta),
\end{equation}
using a trigonometric difference formula.  Fits of Equation (15) are found for concentric rings of data.  The fitted parameters are,
\begin{equation}
I_{\footnotesize 2x}(r) = I_{\footnotesize 2}(r)\mbox{cos}(2\phi_{I2}(r)),
\end{equation}
and,
\begin{equation}
I_{\footnotesize 2y}(r) = I_{\footnotesize 2}(r)\mbox{sin}(2\phi_{I2}(r)).
\end{equation}
The estimated location of the potential minimum is found by calculating,
\begin{equation}
\phi_{I2}(r) = \frac{1}{2}\mbox{tan}^{\mbox{-}1}\frac{I_{\footnotesize 2y}(r)}{I_{\footnotesize 2x}(r)}.
\end{equation}

\section{MODEL FITTING METHODS}

\subsection{Parameter Estimation}

The model parameters are fit to the data using generalized inverse methods.  The following explanation is summarized from Aster et al. (2005, Chapters 4 and 5).  Consider the general form for a linear system of equations, 
\begin{equation}
{\boldsymbol d} = {\boldsymbol G}{\boldsymbol \beta}.
\end{equation}
The column matrix of data, ${\boldsymbol d}$, represents $V_y$ in Equation (3); $\mathcal{V}_y$ in Equations (8), (10), and (11); or $I$ in Equation (15).  The right-hand side includes the operator matrix, ${\boldsymbol G}$, and the column matrix of fitted parameters, $\boldsymbol \beta$.  Let the length of $\boldsymbol \beta$ be $p$ and the length of $\boldsymbol d$ be $n$, then ${\boldsymbol G}$ has $p$ columns and $n$ rows.  The equations for the velocity field and $\Omega_p$ models are weighted using the uncertainties for $\boldsymbol d$. 

The generalized inverse solution is found by performing the singular value decomposition ${\boldsymbol G}$ = ${\boldsymbol U}{\boldsymbol S}{\boldsymbol V^T}$, and then calculating,
\begin{equation}
{\boldsymbol \beta} = {\boldsymbol V}^{-T}({\boldsymbol S}^T{\boldsymbol S})^{-1}{\boldsymbol S}^T{\boldsymbol U}^T{\boldsymbol d}.
\end{equation}
The superscript $T$ indicates matrix transpose.  The columns of ${\boldsymbol U}$ are unit basis vectors spanning the data space.  The diagonal matrix ${\boldsymbol S}$ contains the singular values (note that this is different than the integrated source function, $\mathcal{S}$).  The columns of ${\boldsymbol V}$ are unit basis vectors spanning the model parameter space. 

Regularized fits of the general TW84 method are found using Tikhonov regularization (Aster et al. 2005, Chapter 5).  Tikhonov regularization minimizes,
\begin{equation}
\hskip 8pt||{\boldsymbol d} - {\boldsymbol G}{\boldsymbol \beta}||^2_2 + \lambda^2||{\boldsymbol D}{\boldsymbol \beta}||^2_2,
 \end{equation}
where ${\boldsymbol D}$ is the identity matrix for zeroth-order regularization, or a finite difference operator for higher-order regularization.  The amount of regularization is determined by the value of $\lambda$.  The generalized inverse solution is found by performing an additional singular value decomposition ${\boldsymbol D}$ = ${\boldsymbol W}{\boldsymbol M}{\boldsymbol V^T}$, and then calculating,
\begin{equation}
{\boldsymbol \beta} = {\boldsymbol V}^{-T}({\boldsymbol S}^T{\boldsymbol S}+\lambda^2{\boldsymbol M}^T{\boldsymbol M})^{-1}{\boldsymbol S}^T{\boldsymbol U}^T{\boldsymbol d}.
\end{equation}
The diagonal matrix ${\boldsymbol M}$ contains the singular values for the decomposition of ${\boldsymbol D}$.

The value of $\lambda$ is chosen using the L-curve criteria.   The L-curve criteria finds a compromise between minimizing the residual norm, $||{\boldsymbol G}{\boldsymbol \beta} - {\boldsymbol d}||_2$, and the bias of smoothing from minimizing the solution norm, $||{\boldsymbol D}{\boldsymbol \beta}||_2$.  Plots of $||{\boldsymbol D}{\boldsymbol \beta}||_2$ as a function of $||{\boldsymbol G}{\boldsymbol \beta} - {\boldsymbol d}||_2$ on a log-log scale for a range of $\lambda$ values typically show an L shape.  The L-curve criteria adopts the value of $\lambda$ corresponding to the corner of the L shape.   

\subsection{Uncertainties for the Results}

The uncertainties for the results are reported as 95\% confidence intervals (CIs).  The 95\% CIs are calculated by multiplying the standard errors (SEs) by the appropriate value of $t$ from the Student's $t$-distribution (Ramsey \& Schafer 2012, Chapter 2).  The SEs for the fitted parameters are estimated using the jackknife method (Feigelson \& Babu 2012, Chapter 3).  The SEs for the calculated parameters are found by propagating the SEs for the fitted parameters through the calculation.   

It is common practice to take into account an incorrect value for $\phi_p$ when estimating the uncertainty for $\Omega_p$.  Debattista (2003) and Debattista \& Williams (2004) demonstrate that the effect is non-negligible when using data from long-slit spectroscopy.  The effect is excluded in this paper for three reasons.  The first reason is that the adopted value of $\phi_p$ for NGC 1365 is consistent with most other measurements, and is therefore assumed to be well known (Section 3.1).  The second is that, compared with the errors in $\Omega_p$ from an incorrect value for $\phi_p$ reported by Debattista (2003) and Debattista \& Williams (2004), the errors are typically smaller ($\sim$1/4), and are well within the 95\% CIs, for the high quality velocity field data used in this paper.  The third is that a probability based interpretation of the effect is nontrivial.  Monte Carlo uncertainty estimates that vary the value of $\phi_p$ produce distributions for $\Omega_p$ that differ significantly from common distributions such as a gaussian.  Examples of the second and third reasons are provided in SW11, and Speights \& Westpfahl (2012).  

\subsection{Hypothesis Testing}

Two $t$-tests are included with the results for the $\Omega_p$ models.  One is a test of the null hypothesis that $S$ = 0.  The other is a test of the null hypothesis that $\triangle\Omega_p$ = 0.  For a parameter $\beta$, a test of the null hypothesis $\beta$ = 0 is performed by calculating $t$ = $\beta$/SE, and then finding the probability, $P_{\beta=0}$, for obtaining a value of $t$ should the null hypothesis be true (Ramsey \& Schafer 2012, Chapter 2).  A two-sided $P_{\beta=0}$ value is reported to account for both positive and negative values of $S$ or $\triangle\Omega_p$.  This paper adopts the convention that a value of $P_{\beta=0}$ $<$ 1\% is convincing evidence that $\beta$ $\neq$ 0, a value of 1\% $\leqslant$ $P_{\beta=0}$ $\leqslant$ 5\% is suggestive, and a value of $P_{\beta=0}$ $>$ 5\% is inconclusive.  A lower limit of 0.01\% is adopted for reporting values of $P_{\beta=0}$.

\subsection{Model Comparison}

The models are compared using Bayesian information criterion,
\begin{equation}
\mbox{BIC} = 2\,\mbox{ln}\{\mathcal{L}\} - p\,\mbox{ln}\{n\},
\end{equation}
(Feigelson \& Babu 2012, Chapter 3).  This combination of the likelihood function, $\mathcal{L}$, and the penalty for including $p$ parameters given $n$ data points, provide a balance between underfitting and overfitting the data.  The data uncertainties are expected to approximately follow a gaussian distribution.  When this distribution is used for the likelihood function,
\begin{equation}
\mbox{BIC} =\sum_{j = 1}^{n}\frac{1}{2\pi\sigma_j^2}-\sum_{j = 1}^{n}\frac{({\boldsymbol G_{j,\cdot}}{\boldsymbol \beta} -  d_j)^2}{\sigma_j^2}- p\,\mbox{ln}\{n\}.
\end{equation}
In Equation (24), $\sigma_j$ is the uncertainty for $d_j$.  The $j,\cdot$ notation in ${\boldsymbol G_{j,\cdot}}$ indicates all columns in the $j$th row. The first term on the right-hand side of Equation (24) is the same for the comparison of a set of models, so it is excluded from the reported values.  This paper adopts the convention that an increase in BIC of at least 10 is convincing evidence for a better fit to the data (Kass \& Raftery 1995).  

Equation 24 is derived by approximating the logarithm of Bayes' theorem,
\begin{equation}
P({\boldsymbol \beta } |{\boldsymbol d}) = \frac{P({\boldsymbol d } |{\boldsymbol \beta}) \, P({\boldsymbol \beta}) }{P({\boldsymbol d}) },
\end{equation}
(Schwarz 1978). In Equation (25), $P({\boldsymbol \beta } |{\boldsymbol d})$ is the probability of the model given the data, $P({\boldsymbol d } |{\boldsymbol \beta})$ is the probability of obtaining the data given the model, $P({\boldsymbol \beta})$ is the prior evidence for the model, and $P({\boldsymbol d}$) is the probability of obtaining the data. Bayes' theorem is a way to take into consideration additional, prior knowledge, when calculating a probability.

BIC is ideally suited for the purpose of this paper because it is capable of comparing the non-nested $\Omega_p$ models. (An example of nested model selection is when determining if a trend in the data is best described by a sloping line or a constant line.) The popular $F$ and $\chi^2$ tests (Ramsey \& Schafer 2002; chapters 13 and 19, respectively), for example, require that the models are nested. It is also worth noting that BIC is capable of providing much more compelling evidence for multiple values of $\Omega_p$ than the other widely used Akaike information criterion (Akaike 1974) because BIC provides a much larger penalty for too many parameters in the model. However, it was found that when Akaike information criterion is used instead of BIC, the conclusions of this paper are the same, so only values of BIC are reported.

\subsection{Uncertainties for the Data}

Uncertainties for ${\boldsymbol d}$ are used for weighting the equations for the velocity field and $\Omega_p$ models, and for calculating the BICs.  For the velocity field model, the uncertainties are estimated by Equation (2).  For the $\Omega_p$ models, the uncertainties are estimated by propagating the uncertainties in $V_y$ and $I$ through the calculation of $\mathcal{V}_y$.  
  
\subsection{Correlations in the Data}

The methods above assume the data are uncorrelated.  The correlations are accounted for by replacing the number of data points, $n$, in calculations of the 95\% CIs and the BICs by a reduced number of data points, $n_{\mbox{\footnotesize r}}$ $=$ $n/n_{\mbox{\footnotesize c}}$, for  $n_{\mbox{\footnotesize c}}$ correlated data points.  For the 95\% CIs, the value of $n$ is replaced by $n_r$ in the calculation of the degrees of freedom when finding the value of $t$ from a Student's $t$ distribution. For the BICs, the value of $n$ is replaced by $n_r$ in Equation (24).

The H$\alpha$ data are correlated on a scale that ranges from 2.4 $\times$ 2.4 pixels (due to atmospheric seeing) to 5 $\times$ 5 pixels (due to binning for low signal-to-noise).  The larger correlation size of 5 $\times$ 5 pixels is adopted for this data. The velocity field model samples the H$\alpha$ data in an area because the model is fit to rings having a width of 13$^{\prime\prime}$ (Section 5.1). The number of correlated pixels for the velocity field model is therefore approximated from the number of correlated pixels in an area, $n_c$ = 5$^2$ = 25. The $\Omega_p$ models sample the H$\alpha$ data in adjacent integration paths evaluated at different $y_o$ (Section 5.2). Each integration is therefore correlated with its 4 nearest neighbors, adding up to $n_c$ = 5 correlated integrals.

The K$_{\mbox{\footnotesize s}}$ band data are correlated on a scale of 2.5 $\times$ 2.5 pixels due to atmospheric seeing.  The potential minimum model samples the K$_{\mbox{\footnotesize s}}$ band data in an area because the model is fit to rings having a width of 6\farcs 63 (Section 5.3). For the same reasoning used for the velocity field model, the number of correlated pixels for the potential minimum model is therefore approximated from the number of correlated pixels in an area, $n_c$ = 2.5$^2$ = 6.3. 

\section{RESULTS}

\subsection{Results for the Velocity Field Model}

Equation (3) is fit to the velocity field data using 13$^{\prime\prime}$ wide rings.  This corresponds to approximately twice the number of correlated pixels measured parallel to the kinematic minor ($y$) axis, and projected to the plane of the galaxy disk.  It balances larger scatter and 95\% CIs from using smaller rings with poor angular resolution in the radial behavior of the results from using larger rings.  The warp is avoided by restricting the fit to $r$ $\leqslant$ 220$^{\prime\prime}$. 

Figure 2 shows the radial profile of $V_{\footnotesize \theta}$.  Included in the figure is a profile made by cubic spline interpolating $V_{\footnotesize \theta}$ in increments of 1$^{\prime\prime}$.  The interpolated profile is used in Section 5.2 for identifying possible locations of CR, ILR, and OLR in the plots of the results for $\Omega_p$. 

\begin{figure}
\begin{centering}
\includegraphics[width=0.445\textwidth]{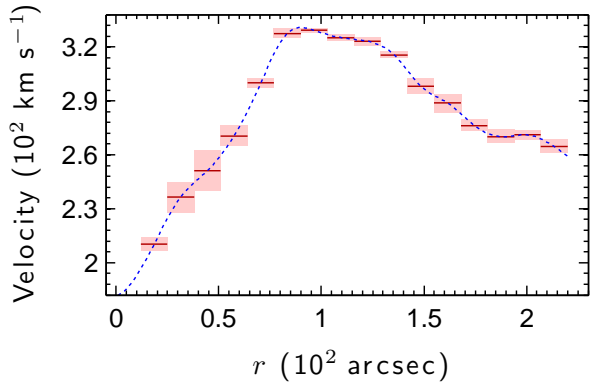} 
\caption{Radial profile of $V_{\theta}$.  The solid red line segments show the fitted results.  The light red shading shows the regions within the 95\% CIs for $V_{\theta}$. The dotted line is interpolated from the fitted results.}
\end{centering}
\end{figure}

Figure 3 shows a map of the modeled velocity field and a map of the residuals.  The largest residuals occur near the bar, and along the concave edge of the spiral pattern.  Near the bar they are as large as $\sim$150 km s$^{-1}$, and are consistent with elliptical streaming.  Near the concave edge of the spiral they are as large as $\sim$50 km s$^{-1}$.  Noncircular motions of this magnitude are also reported by Tueben et al. (1986, hereafter T86) and JM95. It is these noncircular motions that are detected by the TW84 method. In Section 6.2 it is shown that the measured values of $\Omega_p$ for the bar are consistent with the large residuals for the modeled velocity field near the bar.

The patterns characteristic of incorrect values for $\phi_i$, $\phi_p$, $V_{\mbox{\footnotesize sys}}$, or the kinematic center (e.g., van der Kruit \& Allen 1978, Figure 1) are absent from the map of the residuals.  For example, an incorrect value for $\phi_i$ will result in an overabundance of positive residuals on one side of the kinematic minor ($y$) axis and along the major ($x$) axis, and an overabundance of negative residuals on the other side of the kinematic minor axis and along the major axis. An incorrect value for $\phi_p$ will result in an overabundance of positive residuals on one side of the kinematic major axis, and an overabundance of negative residuals on the other side. 

\begin{figure}
\begin{centering}
\includegraphics[width=0.445\textwidth]{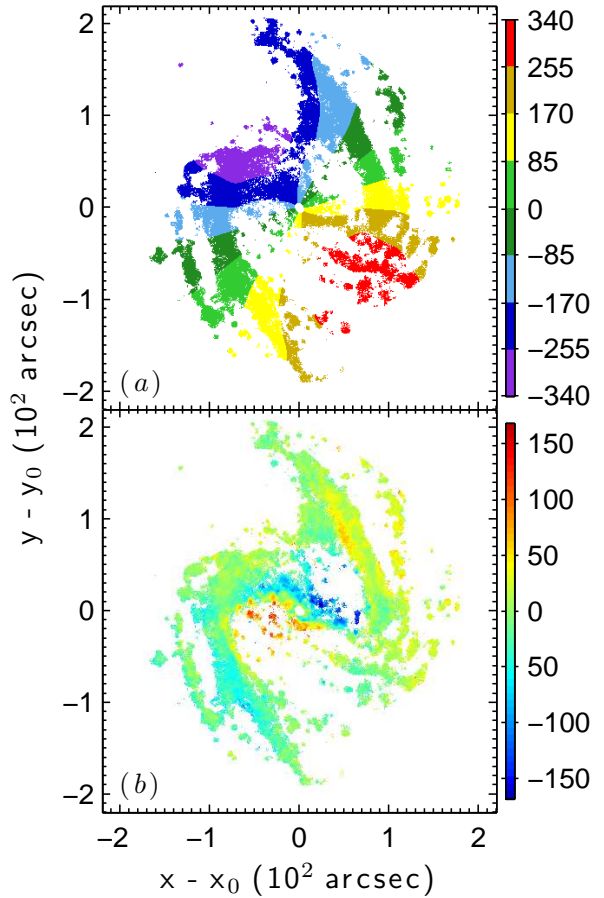} 
\caption{Maps of the velocity field model (panel ($a$)) and the residuals (panel ($b$)).  The horizontal and vertical axes are scaled in the same way as Figure 1.  The map of the velocity field model is binned in increments of 85 km s$^{-1}$.  The colorbars are in units of km s$^{-1}$.}
\end{centering}
\end{figure}

\subsection{Results for the $\mathit \Omega_p$ Models}

Equations (8), (10), and (11) are fit to the H$\alpha$ data by first rotating the maps to align the pixel gridding parallel to the $x$ and $y$ axes. The rotated maps have the same pixel size of 0\farcs 98 as the originals. Maps of the integrands are then calculated from the rotated maps. Integration is performed by summing the values for the pixels along the $x$ direction in the maps of the integrands.

The locations of the integration paths are limited to $|y_o|$ $<$ 220$^{\prime\prime}$.  This results in 86\% of the integrands converging before the beginning of the warp at $r$ $=$ 240$^{\prime\prime}$.  All of the integrands converge by $r$ $=$ 274$^{\prime\prime}$, where the value of $\phi_p$ is $\approx$ 3$^{\circ}$ smaller than the adopted value of 220$^{\circ}$ (JM95, Figure 11).  The effect of the integrals that converge at $r$ $\geqslant$ 240$^{\prime\prime}$ are determined to be negligible by comparing the results from applying the general TW84 method with and without the outermost ring (following the procedure in Section 5.2.2).  The results for both applications differ by an amount that is much smaller than the 95\% CIs. 

The results are plotted for $r$ $\leqslant$ 220$^{\prime\prime}$.  Plotting the results for larger $r$ reduces the resolution of the results along that axis, which is unnecessary for the purpose of this paper.  The transition from the bar to the spiral pattern occurs about halfway across the plots.

Observational units of km s$^{-1}$ arcsec$^{-1}$ are used for reporting angular frequencies.  This avoids the need to assume a distance to NGC 1365 for the conversion to galactic units of km s$^{-1}$ kpc$^{-1}$, and simplifies comparisons with previous results.  For the interested reader, the average of 43 distances found in the NASA/IPAC Extragalactic Database is 18.2 Mpc.  Assuming this distance, the conversion factor is 1$^{\prime\prime}$ = 0.09 kpc. 

\begin{deluxetable*}{lllcl}[b!]
\tablecaption{Summary of the $\Omega_p$ Model Fits}
\tablewidth{0pt}
\startdata
\tableline\tableline
Fit & Method &  Regularization & BIC & Notes\\
\tableline
1	& Original & None & -3085  & Assumes a single, global value of $\Omega_p$ \\
2	& Original & None & \hskip 5.5pt -635 & Fits for different $\Omega_p$ in different radial regions\\
3	& General & None & \hskip 5.5pt -620 & Rings are adopted from the regions in Fit 2 \\
4	& General & None & \hskip 5.5pt -609 & Ring width = 36\farcs 7 \\
5	& General & None & \hskip 5.5pt -387 & Ring width = 10$^{\prime\prime}$ \\
6	& General & Zeroth order & -1606 & Ring width = 10$^{\prime\prime}$ \\
7	& General & First order & -1261 & Ring width = 10$^{\prime\prime}$ \\
8	& General &	Second order & \hskip 5.5pt -930 & Ring width = 10$^{\prime\prime}$ \\
9	& General &	First order & \hskip 5.5pt -696 & Ring width = 10$^{\prime\prime}$, excludes $|y_o|$ $<$ 10$^{\prime\prime}$\\
10	& General & None & \hskip 5.5pt -375 & Uses BIC to select model discontinuities  \\
11	& General & None & \hskip 5.5pt -375 & Uses BIC to select model discontinuities  
\enddata 
\end{deluxetable*}

The results for 11 model fits are presented, hereafter referred to as Fit 1 -- Fit 11.  They differ in the form of the TW84 method, whether the fits are regularized or not, and the radial ranges or ring widths.  Basic properties of the fits are summarized in Table 2. 

\subsubsection{Results for the Original TW84 Method}

Figure 4 shows lines fit to $\mathcal{V}_y(\mathcal{X})$ using Equation (8).  They are Fit 1 and Fit 2 in Table 2.  The slopes of the lines are $\Omega_p$ in Table 3.   Panel ($a$) at the top of the figure shows Fit 1 for a single value of $\Omega_p$.  There are, however, three different and clearly defined linear trends in $\mathcal{V}_y(\mathcal{X})$ that are inconsistent with a single value of $\Omega_p$.   Panels ($b$), ($c$), and ($d$) show what is collectively Fit 2 for the three trends.  They occur in ranges of density weighted radii,  
\begin{equation}
r_w=\frac{\int^{\infty}_{-\infty}I(x,y_o)r(x,y_o)\,dx}{\int^{\infty}_{-\infty}I(x,y_o)\,dx},
\end{equation}
that are easily identifiable from a combination of trial-and-error and visual inspection.  The ranges of $r_w$ are indicated in the top of panels ($b$), ($c$), and ($d$).  The increase in BIC from Fit 1 to Fit 2 is 2450.  This is convincing evidence that the three different trends are real.  

\begin{figure}[t!]
\begin{centering}
\includegraphics[width=0.38\textwidth]{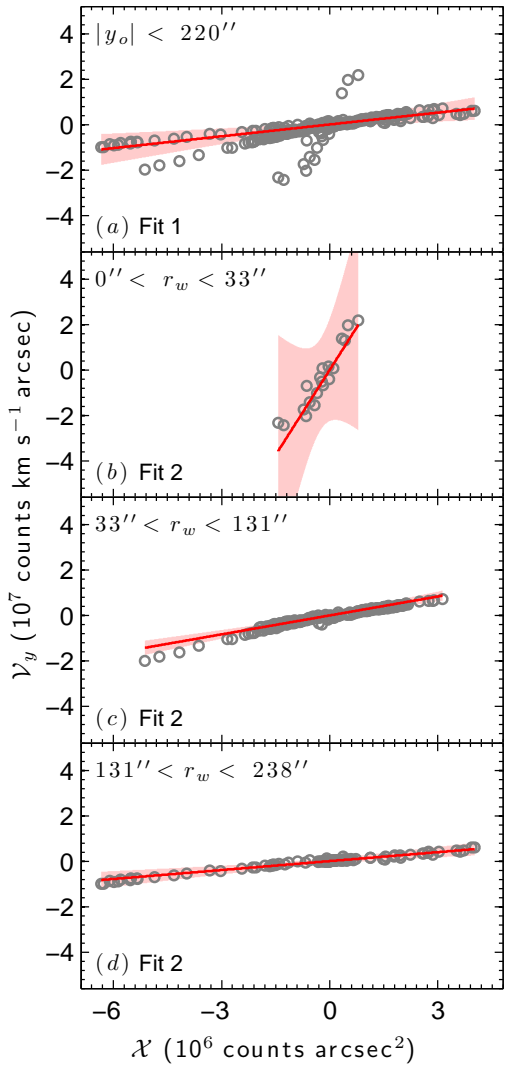} 
\caption{Fitted lines for the original TW84 method. The gray circles show the calculated values of $\mathcal{V}_y(\mathcal{X})$.  The red solid lines show the model fits.  The light red shading shows the region within the 95\% confidence bands for a fitted line.  The fit numbers and regions are indicated in the panels.}
\end{centering}
\end{figure}

The detection of a different value of $\Omega_p$ for the innermost range in $r_w$ is consistent with previous findings of a different pattern in the nuclear region (Jungwiert et al. 1997, Emsellem et al, 2001, Laine et al. 2002).  A detailed analysis of $\Omega_p$ for the nuclear region is beyond the scope of this paper.  Similar results in the remainder of this paper are therefore merely reported.

Figure 5 shows radial profiles of $\Omega_p$ for Fit 1 and Fit 2.  Included in the figure are possible locations of CR, ILR, and OLR.  The result for the innermost region of Fit 2 is excluded so that the scale of the angular frequency axis is small enough to easily see the 95\% CIs for the other results.   

\begin{figure}
\begin{centering}
\includegraphics[width=0.38\textwidth]{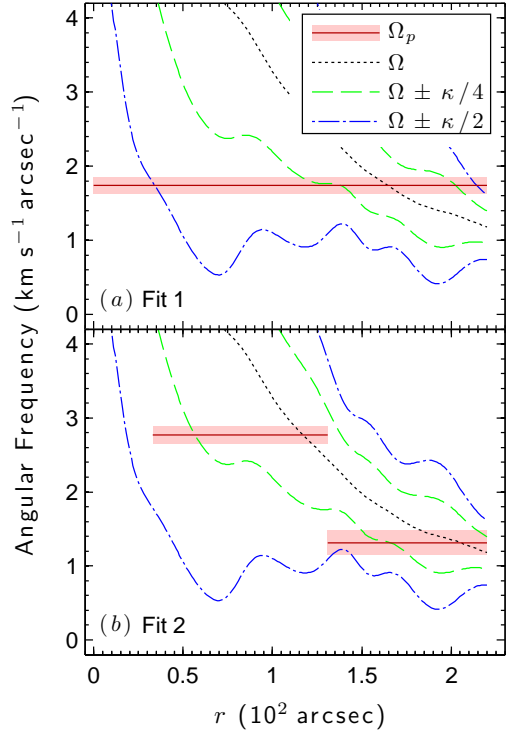} 
\caption{Radial profiles of $\Omega_p$ for the original TW84 method.  A legend is provided in panel ($a$).  The light red shading shows the regions within the 95\% CIs for $\Omega_p$.  The fit number is indicated in the panels.}
\end{centering}
\end{figure}

\begin{deluxetable*}{lccccc}
\tablecaption{Pattern Speed Results for the Original Method}
\tablewidth{0pt}
\startdata
\tableline\tableline
 Fit & Region & $\Omega_p$ & $\Omega^{\prime}_p$ & $\triangle$$\Omega_p$ & $P_{\triangle\Omega_p=0}$ \\
& (arcsec) & (km s$^{-1}$ arcsec$^{-1}$) & (km s$^{-1}$ arcsec$^{-1}$) & (\%) & (\%) \\
\tableline
1 &  -220 $<$ \hskip 0.5pt $y_o$ \hskip -0.5pt $<$ 220 & \hskip 4pt 1.74 $\pm$ \hskip 4pt 0.11 & \hskip 4pt 1.75 $\pm$ \hskip 4pt 0.11 & \hskip 2.5pt -0.62 $\pm$ \hskip 8.5pt 8.70 & 96.41 \\
2 & \hskip 7pt 10 $\leqslant$ \hskip 1pt $r_w$ \hskip -2.5pt $<$ \hskip 4pt 33 & 24.99 $\pm$ 17.05 & 24.02 $\pm$ 14.01 & \hskip 5.5pt 4.06 $\pm$ 119.27 & 87.55 \\
2 & \hskip 7pt 33 $<$ \hskip 1pt $r_w$  \hskip -2pt $<$ 131 & \hskip 4pt  2.77 $\pm$ \hskip 4pt 0.12 & \hskip 4pt 2.68 $\pm$ \hskip 4pt 0.10 & \hskip 5.5pt 3.26 $\pm$ \hskip 8.5pt 5.83 & 70.53\\
2 & \hskip 2.5pt 131 $<$ \hskip 1pt $r_w$ \hskip -1.5pt  $\leqslant$ 237 & \hskip 4pt 1.31 $\pm$ \hskip 4pt 0.05 & \hskip 4pt 1.33 $\pm$ \hskip 4pt 0.05 & \hskip 2.5pt -1.09 $\pm$ \hskip 8.5pt 5.68 & 92.95
\enddata 
\end{deluxetable*}

\begin{deluxetable}{lccc}
\tablecaption{Integrated Source Function Results for the Original  Method}
\tablewidth{0pt}
\startdata
\tableline\tableline
Fit & Region & $S$ & $P_{\footnotesize S=0}$\\
& (arcsec) & (10$^5$ counts km s$^{-1}$ arcsec) & (\%) \\
\tableline
1 & -220 $<$ \hskip 0.5pt $y_o$ \hskip -0.5pt $<$ 220  & \hskip 2.7pt 0.25 $\pm$ \hskip 8.7pt 0.20 & \hskip 9.3pt 1.37\\
2 & \hskip 7pt 10 $\leqslant$ \hskip 1pt $r_w$ \hskip -2.5pt $<$ \hskip 4pt 33 & -6.57 $\pm$ 131.47 & \hskip 5pt 88.51\\
2 & \hskip 7pt 33 $<$ \hskip 1pt $r_w$  \hskip -2pt $<$ 131 & \hskip 0pt -4.33 $\pm$ \hskip 8.7 pt 0.82 & $<$ 0.01 \\
2 &  \hskip 2.5pt 131 $<$ \hskip 1pt $r_w$ \hskip -1.5pt  $\leqslant$ 237& \hskip 2.7pt 0.33 $\pm$ \hskip 8.7pt 0.16 & \hskip 9.3pt 0.04 
\enddata 
\end{deluxetable}

Most of the calculated values of $\mathcal{V}_y(\mathcal{X})$ in Figure 4 are well described by the straight lines in panels ($b$), ($c$), and ($d$). The effect of violating assumption (3) is checked by fitting Equation (11) with a constant value for $\mathcal{S}$.  The results for Equation (11) are found to be consistent with those shown in Figures 4 and 5.  They show three different linear trends that are clearly defined.  The ranges of $r_w$ are also the same.  The increase in BIC for a fit with a single value of $\Omega_p^\prime$ to one with the three trends is 2549.  

Table 4 shows the values of $S$ and $P_{S=0}$ for fits of Equation (11).  The value of $P_{S=0}$ for $|y_o|$ $<$ 220$^{\prime\prime}$ is suggestive, but inconclusive, for ruling out the null hypothesis that $S$ = 0.  The value of $P_{S=0}$ for 10$^{\prime\prime}$ $\leqslant$ $r_w$ $<$ 33$^{\prime\prime}$ is too large to rule out the null hypothesis. The values of $P_{S=0}$ for 33$^{\prime\prime}$ $<$ $r_w$ $<$ 131$^{\prime\prime}$ and 131$^{\prime\prime}$ $<$ $r_w$ $\leqslant$ 237$^{\prime\prime}$ are small enough to rule out the null hypothesis.    

Although there is evidence for nonzero values of $S$ for 33$^{\prime\prime}$ $<$ $r_w$ $<$ 131$^{\prime\prime}$ and 130$^{\prime\prime}$ $<$ $r_w$ $\leqslant$ 237$^{\prime\prime}$, excluding $S$ has a negligible effect on the values of $\Omega_p$.  Included in Table 3 are values of $\Omega_p^{\prime}$, $\triangle\Omega_p$, and $P_{\triangle\Omega_p=0}$.  The values of $\triangle\Omega_p$ are indistinguishable from 0 given the size of their 95\% CIs.  The values of $P_{\triangle\Omega_p=0}$ are too large to rule out the null hypothesis that $\triangle\Omega_p$ = 0.  The negligible effect of excluding constant values for $S$ in the model fits is explainable by noting that it is 2--3 orders of magnitude smaller than most values of $\mathcal{V}_y$.      

The bar and beginning of the spiral pattern appear to share a common value $\Omega_p$, but this result is unreliable due to violating assumption (2) of the original TW84 method. Consider that integration paths in the bar region will also cross some of the spiral pattern. If the spiral pattern is rotating more slowly than the bar, $\Omega_p$ will be underestimated in the bar region.  Integration paths with $r_w$ outside the bar region may still cross some of the bar, and this can result in an overestimate of $\Omega_p$ for the beginning of the spiral pattern. The combination of these two effects can result in what appears to be a single slope in panel (c) of Figure 4. Furthermore, although the data are well described by a straight line in that panel, it is imperfect. The results for Fit 1, and the larger values of $\Omega_p$ for the bar found using the general TW84 method (Section 5.2.2.) are consistent with this interpretation. The division of Fit 2 into three trends most importantly demonstrates that assumption (2) is violated.

\subsubsection{Results for the General TW84 Method}

The results for the original TW84 method show that assumptions  (2) and (3) of the method are violated.  Although violations of assumption (3) are approximately negligible, the violations of assumption (2) are large.  The radial profile of $\Omega_p$ is therefore found using the general TW84 method.  It is assumed that violations of assumption (3) also have a negligible effect on the results for the general TW84 method.

Figure 6 shows radial profiles of $\Omega_p$ for unregularized fits of Equation (10).  They are Fit 3 -- Fit 5 in Table 2.  The inner and outer radii of the rings for Fit 3 are equivalent to the lower and upper limits, respectively, of the regions for Fit 2, with the addition of an outermost ring from $r$ $>$ 237$^{\prime\prime}$ to the edge of the data. Fit 4 has a ring width of $\approx$ 36.7$^{\prime\prime}$ for $r$ $\leqslant$ 220$^{\prime\prime}$. This ring width is calculated by dividing 220$^{\prime\prime}$ into six equal parts, and is chosen because it provides a discontinuity in the model near the end of the bar pattern. Fit 5 has a ring width of 10$^{\prime\prime}$ for $r$ $\leqslant$ 220$^{\prime\prime}$. The outermost rings for Fit 4 and Fit 5 start at $r$ = 220$^{\prime\prime}$, and extend to the edge of the data.

\begin{figure*}[!t]
\begin{centering}
\includegraphics[width=1\textwidth]{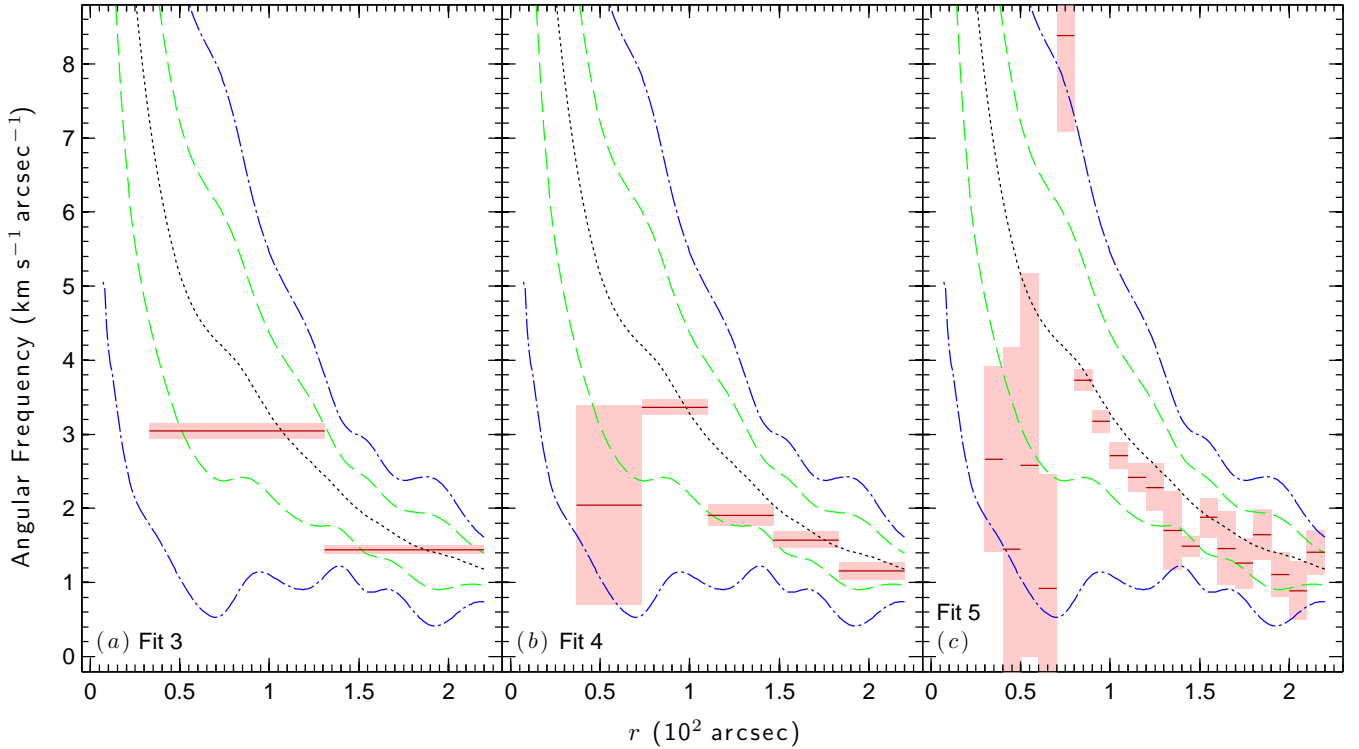} 
\caption{Radial profiles of $\Omega_p$ for the general TW84 method.  The figure is formatted in the same way as in Figure 5.}
\end{centering}
\end{figure*}

Some of the results for the innermost rings are excluded from Figure 6 because including them unnecessarily increases the scale of the angular frequency axes.  The result not shown for Fit 3 is 14.99 $\pm$ 4.51 km s$^{-1}$ arcsec$^{-1}$.  The result not shown for Fit 4 is 12.81 $\pm$ 3.85 km s$^{-1}$ arcsec$^{-1}$.  The results not shown for Fit 5 are 30.87 $\pm$ 5.63, 23.97 $\pm$ 4.36, and 12.1 $\pm$ 1.28 km s$^{-1}$ arcsec$^{-1}$, in order of increasing $r$.   

The results for Fit 3 are similar to those for Fit 2.  The 95\% CIs for $\Omega_p$ overlap in the innermost region and ring in Fit 2 and Fit 3, respectively.  The other two values of $\Omega_p$ are $\sim$10\% larger for Fit 3.  The increase in BIC from Fit 2 to Fit 3 is 15.  This is convincing evidence that Fit 3 is a better fit to the data than Fit 2.    

Fit 4 shows more complex behavior in the radial profile of $\Omega_p$.  The 95\% CIs for the 2 innermost rings of Fit 4 are large.  The values of $\Omega_p$ show a trend that decreases with increasing radius near the end of the bar and in the spiral pattern region.  The increase in BIC from Fit 3 to Fit 4 is 11.  This is convincing evidence that Fit 4 is a better fit to the data than Fit 3. 

The increase in BIC from Fit 4 to Fit 5 is 222, but the solution for Fit 5 is clearly unstable.  When the solution is unstable the value of BIC is unreliable because it can no longer be assumed that the likelihood function has the form of a gaussian distribution. This would change the first and second terms on the right-hand side of Equation (24), the former of which is assumed in this paper to be the same for all models. The 95\% CIs for the innermost rings are large for Fit 5, similar to those for Fit 4.  Despite the increased instability, values of $\Omega_p$ for Fit 5 show a general trend that decreases with increasing radius in the spiral pattern region, similar to the trend for Fit 4. 
  
Figure 7 shows radial profiles of $\Omega_p$ for regularized fits of Equation (10).  They are Fit 6 -- Fit 8 in Table 2.  They use zeroth, first, and second-order regularization, in the order of the fit number.  For $r$ $\leqslant$ 220$^{\prime\prime}$, the ring width is 10$^{\prime\prime}$, equivalent to the ring width for Fit 5.  The outermost rings extend to the edge of the data. 

\begin{figure*}[!hp]
\begin{centering}
\includegraphics[width=1\textwidth]{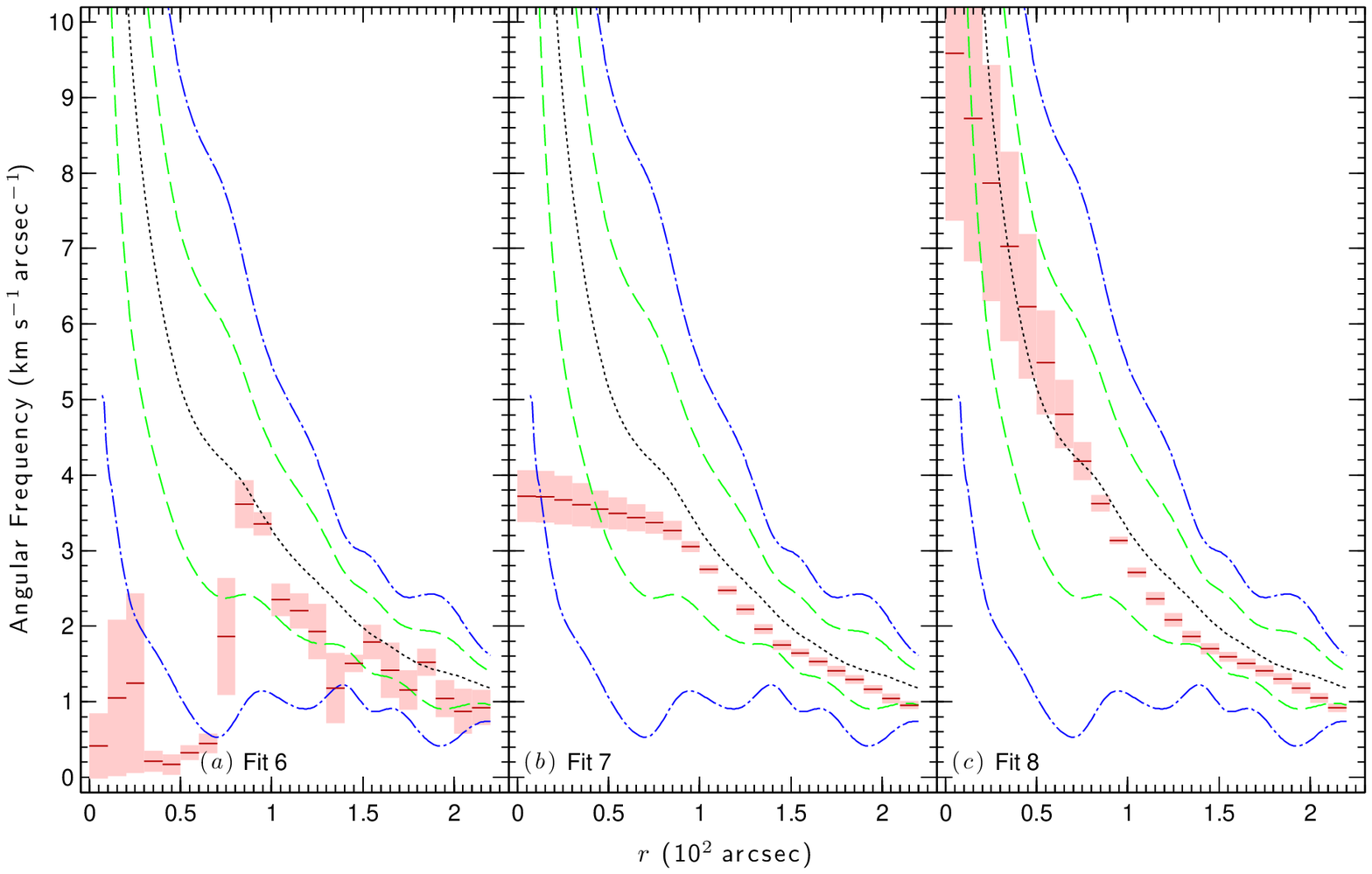} 
\caption{Radial profiles of $\Omega_p$ for regularized solutions of the general TW84 method.  The figure is formatted in the same way as in Figure 5.  The results for zeroth, first, and second-order regularization are shown in panels ($a$), ($b$), and ($c$), respectively.}
\end{centering}
\end{figure*}

Figure 8 shows log-log plots of the solution norm as a function of the residual norm for a range of $\lambda$ values.  The corners of the L-curve for Fit 6, Fit 7, and Fit 8 occur at $\lambda$ = 33.3, 184.2, and 9.2, respectively.  The value of the solution norm at the corner of the L-curve increases for increasing order of regularization, whereas the value of the residual norm decreases for increasing order of regularization.  The corner of the L-curve is best defined for Fit 7.

\begin{figure*}[h]
\begin{centering}
\includegraphics[width=\textwidth]{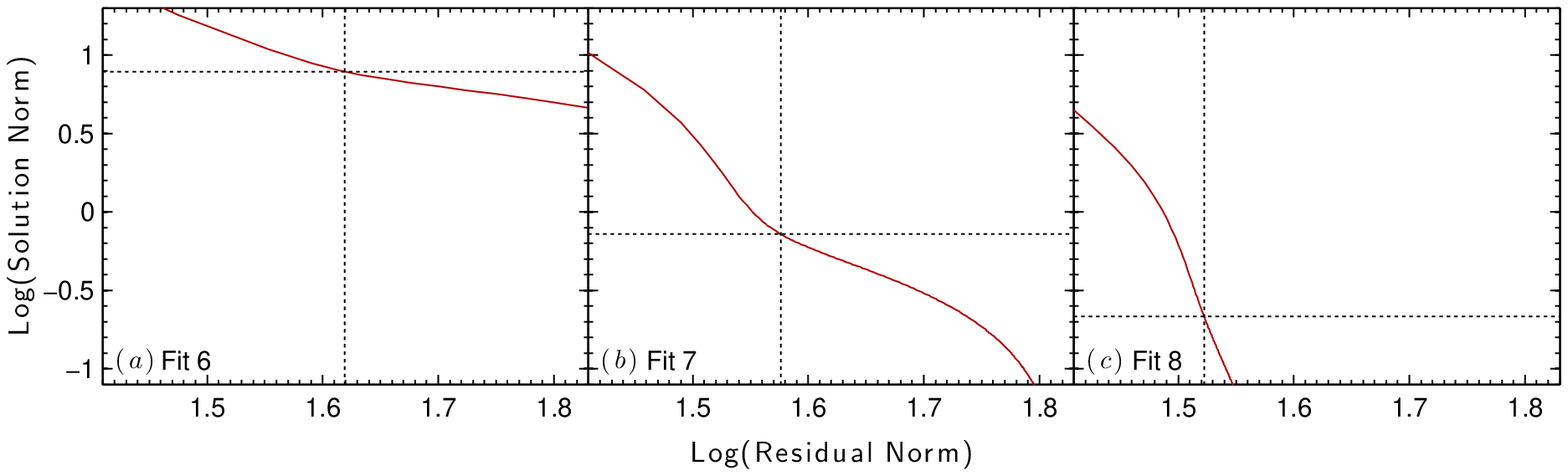} 
\caption{Plots of L-curves used for selecting $\lambda$.  The solid red lines show curves for zeroth, first, and second-order regularization in panels ($a$), ($b$), and ($c$), respectively.  The intersection of the horizontal and vertical dotted lines show the locations of the corners in the L-curve. Each panel shows the same scale in the horizontal and vertical axis for ease of comparison.}
\end{centering}
\end{figure*}

The results for Fit 6 fail to show clearly defined trends for the radial profile of $\Omega_p$.  There are indications of over-regularization in the bar region because the results are biased toward zero for $r$ $\leqslant$ 70$^{\prime\prime}$.  There are indications of under-regularization in the spiral pattern region because the results there are similar to those for Fit 5. 

The results for Fit 7 and Fit 8 show clearly defined trends for the radial profile of $\Omega_p$ that decrease with increasing radius.  The trend in $\Omega_p$ for the spiral pattern region is concave up for both fits.  The results for Fit 7 and Fit 8 differ by less than 10\% in this region, and are indistinguishable for $r$ $\geqslant$ 140$^{\prime\prime}$, given the size of the 95\% CIs. The gradient of the trend for $r$ $\leqslant$ 90$^{\prime\prime}$ is much larger for Fit 8.  The larger gradient is explainable by noting that first-order regularization penalizes gradients, and second-order regularization favors constant gradients.  This is consistent with the existence of an $\Omega_p$ in the nuclear region of the disk that has a larger value than the rest of bar, as found for Fit 2 -- Fit 4.

When integration paths that cross the nuclear region are excluded from the fit, the profile is consistent with an approximately rigid bar.  Figure 9 shows a radial profile of $\Omega_p$ for a first-order regularized fit of Equation (10) that excludes integration paths that are within $|y_o|$ $<$ 10 $^{\prime\prime}$.\linebreak  This is Fit 9 in Table 2.  There is an approximately constant value of $\Omega_p$ for 10$^{\prime\prime}$ $\leqslant$ $r$ $\leqslant$ 90$^{\prime\prime}$.  The radial profile for Fit 9 demonstrates that the TW84 method can detect rigid patterns when allowed to vary with radius.  

\begin{figure}[]
\begin{centering}
\includegraphics[width=0.395\textwidth]{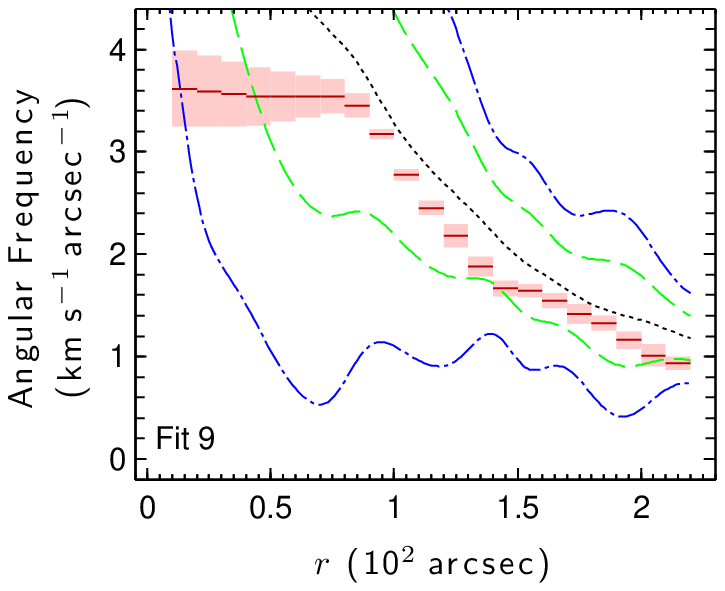} 
\caption{Radial profile of $\Omega_p$ for a regularized solution of the general TW84 method that excludes integration paths crossing the nuclear region.  The figure is formatted in the same way as in Figure 5.}
\end{centering}
\end{figure}

Excluding a larger range in $y_o$ fails to produce evidence for a rigid spiral pattern.  This is determined by excluding successively larger regions in increments of 10$^{\prime\prime}$, $|y_o|$ $<$ 150 $^{\prime\prime}$. The results for these fits are found to be consistent with the results for Fit 9.

The values of BIC for regularized fits are smaller than those for unregularized fits.  The largest value of BIC for the regularized fits is -696, and occurs for Fit 9.  This is adopted as the best fitting model that uses regularization. Differences in the value of BIC that involve regularized fits should be interpreted with caution when comparing models because regularized fits are biased.  The residual norm values in the calculations of BIC, for example, depend on the results for the L-curve criteria.  

\subsubsection{Possible Locations for Discontinuities}  

The results presented in Sections 5.2.1 and 5.2.2 provide evidence for the existence of different values of $\Omega_p$ in different regions of the disk, inconsistent with the bar and spiral patterns having the same value of $\Omega_p$.  To search for evidence of mode coupling, the best possible locations for discontinuities in the model need to be identified, and this is unclear from the results presented so far.  For example, the regularization process favors continuous profiles.  In what follows, possible locations for discontinuities are estimated by fitting models with different radii for discontinuities, and comparing their values of BIC.  They are unregularized in order to avoid that form of bias.  

Figure 10 shows values of BIC for five sets of comparisons.  They are referred to as Set 1 --  Set 5, in the order of comparison.  Each set identifies a radius of discontinuity that is adopted in subsequent sets until there is no longer a significant improvement in the value of BIC.  The possible radii for a discontinuity is changed in 1$^{\prime\prime}$ increments.  The minimum ring width shown is 10$^{\prime\prime}$.  The ranges of possible radii shown are restricted to $r$ $\leqslant$ 220$^{\prime\prime}$, and are labelled in each panel.  

\begin{figure*}
\begin{centering}
\includegraphics[width=1\textwidth]{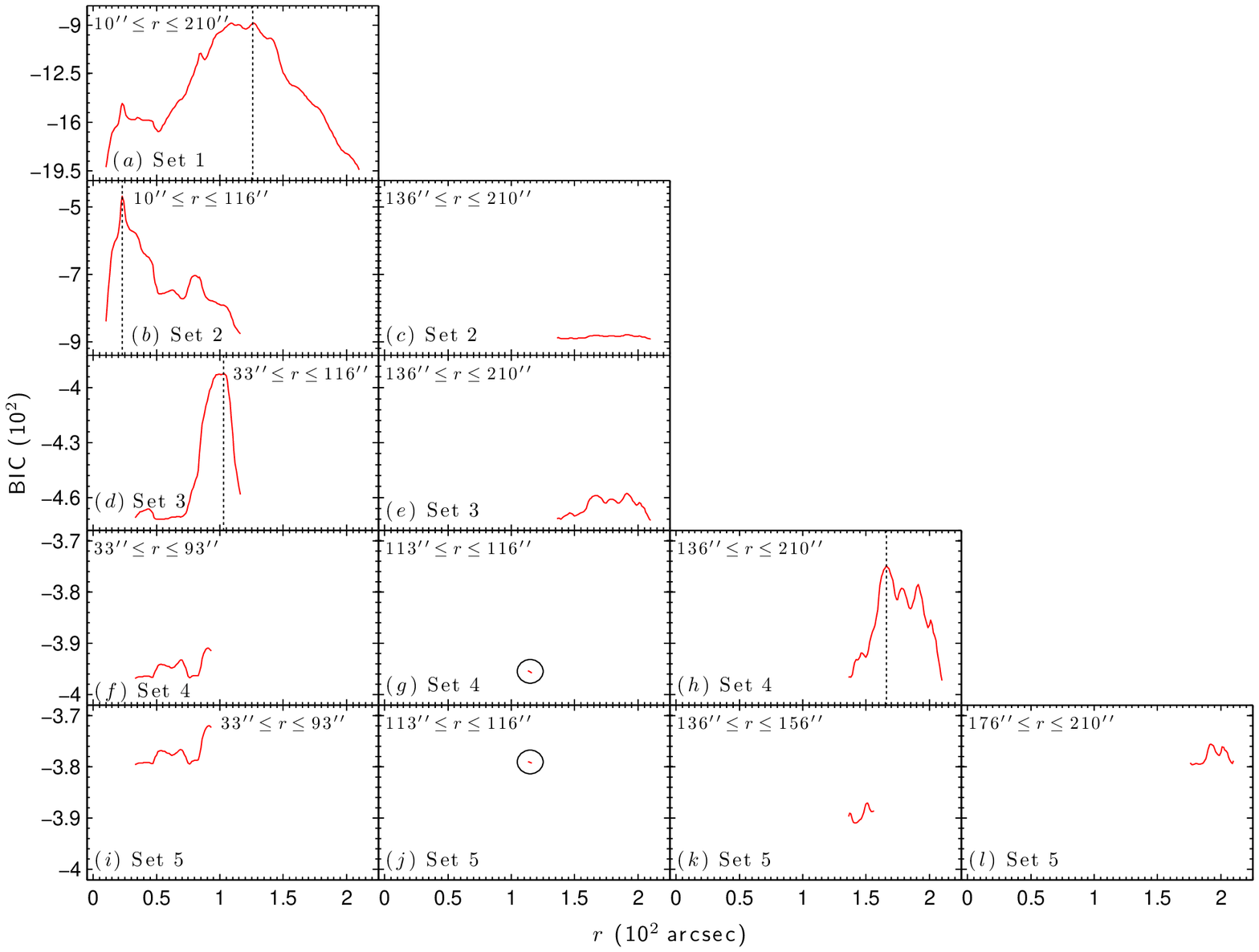} 
\caption{Radial profiles of BIC for choosing the radii for discontinuities in the $\Omega_p$ model.  Each row shows a set of compared BICs.  The vertical dotted lines show the adopted radius for a discontinuity in the $\Omega_p$ model that is used in a subsequent set of compared values.  The radial ranges are indicated in the panels. The BIC profiles in panels ($g$) and ($j$) are circled for easier identification.}
\end{centering}
\end{figure*}

Set 1 is shown in panel ($a$) of Figure 10.  The largest value of BIC is -887, and occurs at $r$ = 126$^{\prime\prime}$.  The increase in BIC from Fit 1 to the largest value in Set 1 is 2198.

Set 2 is shown in panels ($b$) and ($c$).  The largest value of BIC is -468, and occurs at $r$ = 23$^{\prime\prime}$.  The increase in the largest values of BIC from Set 1 to Set 2 is 419.  Subsequent comparisons assume a mean value for $\Omega_p$ for $r$ $<$ 23$^{\prime\prime}$.  Identifying discontinuities in the model for $r$ $<$ 23$^{\prime\prime}$ is unnecessary for the purpose of this paper.

Set 3 is shown in panels ($d$) and ($e$). The largest value of BIC is -392, and occurs at $r$ = 103$^{\prime\prime}$.  The increase in the largest values of BIC from Set 2 to Set 3 is 76.

Set 4 is shown in panels ($f$), ($g$), and ($h$). The largest value of BIC is -375, and occurs at $r$ = 166$^{\prime\prime}$.  The increase in the largest values of BIC from Set 3 to Set 4 is 17.

Set 5 is shown in panels ($i$), ($j$), ($k$), and ($l$).  The largest value of BIC is -372, and occurs at $r$ = 92$^{\prime\prime}$.  The increase in the largest values of BIC from Set 4 to Set 5 is 7.  This is too small to provide convincing evidence for a better fit to the data, so $r$ = 92$^{\prime\prime}$ is not adopted as a radius of discontinuity, thus ending the process.    

The above process produces a list of discontinuities in the model at $r$ = 126$^{\prime\prime}$, 23$^{\prime\prime}$, 103$^{\prime\prime}$, and 166$^{\prime\prime}$, in the order they are identified.  The fitted model using these radii is Fit 10 in Table 2.  It has the largest value of BIC of Fit 1 -- Fit 10 by at least 12.  This is convincing evidence that Fit 10 is a better fit to the data than any of the other fits presented so far.

To check how robust these results are, the above process is repeated starting with the second largest value of BIC in Set 1. It is -888, and occurs at $r$ = 109$^{\prime\prime}$.  Following the same process as above produces a second list of discontinuities at $r$ = 109$^{\prime\prime}$, 23$^{\prime\prime}$, 128$^{\prime\prime}$, 95$^{\prime\prime}$, and 166$^{\prime\prime}$, in the order they are identified.  This is Fit 11 in Table 2.  Choosing a different first radius of discontinuity changes the locations of the discontinuities in the model near the transition from the bar to the spiral pattern.  The same radii for discontinuities are found at $r$ = 23$^{\prime\prime}$ and 166$^{\prime\prime}$.  The value of BIC for Fit 11 is the same as for Fit 10.  

Figure 11 shows radial profiles of $\Omega_p$ for Fit 10 and 11.  The results for the innermost ring are excluded because including them unnecessarily increases the scale of the angular frequency axes.  They are 25.89 $\pm$ 3.44 km s$^{-1}$ arcsec$^{-1}$ for Fit 10, and 25.87 $\pm$ 3.38 km s$^{-1}$ arcsec$^{-1}$ for Fit 11.  The radial profile of $\Omega_p$ is very similar for the two fits.  They both show values of $\Omega_p$ for the bar region that are clearly different than the values for the spiral pattern region, and a general trend that decreases for increasing radius.   The general trends in the radial profiles for Fit 10 and 11 are consistent with the regularized trend in the profile for Fit 9.  In the bar region, most of the results for Fit 10 and 11 are within the 95\% CIs for Fit 9. 

\begin{figure}
\begin{centering}
\includegraphics[width=0.38\textwidth]{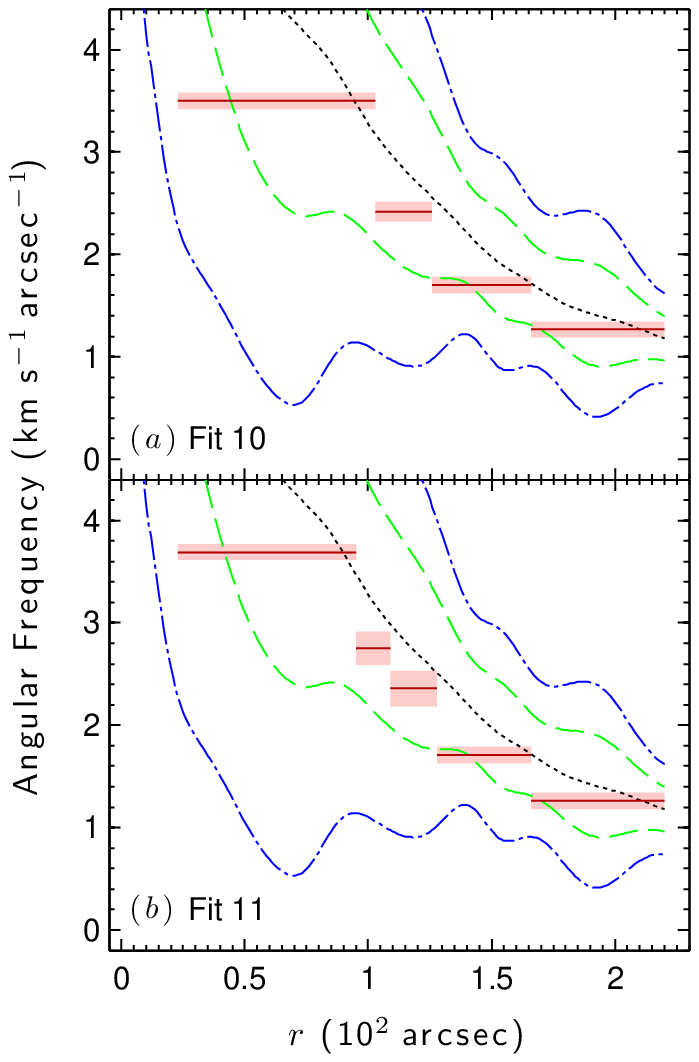} 
\caption{Radial profiles of $\Omega_p$ from using BIC model selection.  The figure is formatted in the same way as in Figure 5.}
\end{centering}
\end{figure}

\subsection{Results for the Potential Minimum Model}

Equation (14) is fit to the K$_{\mbox{\footnotesize s}}$ band data using 6\farcs 63 wide rings.  This is approximately twice the number of correlated pixels measured parallel to the kinematic minor ($y$) axis, and projected to the plane of the galaxy disk.  It is nearly half the resolution of the fit to the velocity field.  The warp is avoided by restricting the fit to $r$ $\leqslant$ 220$^{\prime\prime}$. 

Figure 12 shows the radial profile of $\phi_{I2}$.  The profile is consistent with that shown in Figure 8 of Lindblad et al. (1996) for J band data.  The increasing trend for $r$ $\lesssim$ 34\farcs 4 is due to dust lanes obscuring the starlight in the nuclear region.  For  34\farcs 4 $\lesssim$ $r$ $\lesssim$ 100\farcs 7, $\phi_{I2}$ is approximately constant given the 95\% CIs, consistent with an approximately rigid bar.  For $r$ $\gtrsim$ 100\farcs 7$^{\prime\prime}$, $\phi_{I2}$ increases with increasing radius.  There are a dips in the profile at $r$ = 137$^{\prime\prime}$ and 174$^{\prime\prime}$.  The scatter increases at the latter radius  because the signal-to-noise ratio begins to drop.

A fit of constant $\theta$ to the profile of $\phi_{I2}$ for 34\farcs 4 $\leqslant$ $r$ $\leqslant$ 100\farcs 7 results in a bar position angle of $\phi_b$ = 1.02 $\pm$ 0.03 rad, or 58\fdg 58 $\pm$ 1\fdg 47, measured from $x$ to $y$ in galaxy coordinates.  This is equivalent to 91\fdg 01 $\pm$ 1\fdg 11 degrees measured from north to east in sky coordinates, consistent with Lindblad (1978).  The horizontal dotted line in Figure 12 shows the fit of constant $\theta$ for the position angle of the bar.

The radii of the discontinuities for Fit 10 and 11 are indicated in Figure 12.  The models for both Fit 10 and 11 have discontinuities near the end of the bar, indicating a change in the value of $\Omega_p$ there.  The two dips in the radial profile of $\phi_{I2}$ at $r$ = 177$^{\prime\prime}$ and 137$^{\prime\prime}$ are close to discontinuities in the models for Fit 10 and 11. This could be an indication of slightly offset spiral patterns, or a coincidence due to uncertainties in the results.\\\\

\begin{figure}
\begin{centering}
\includegraphics[width=0.395\textwidth]{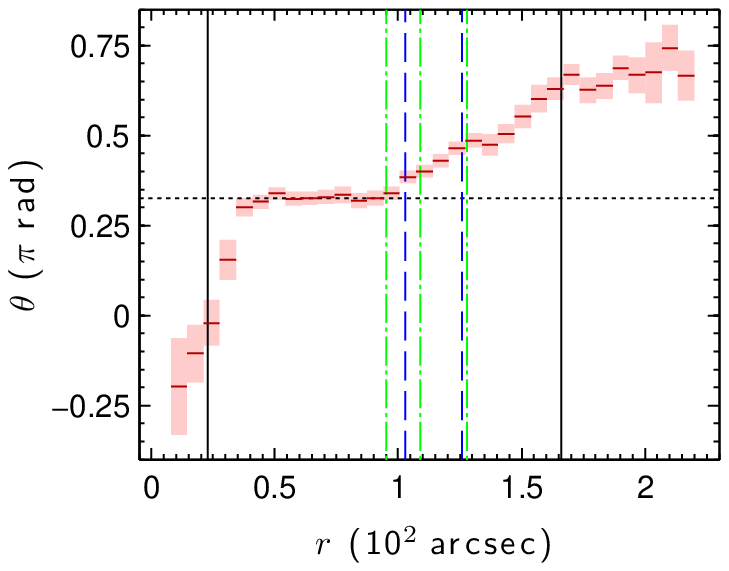} 
\caption{Radial profile of $\phi_{I2}$.  The solid red line segments show the fitted results.  The light red shading shows the regions within the 95\% CIs for $\phi_{I2}$.  The horizontal dotted line shows the fitted bar position angle.  The vertical lines show the radii for discontinuities in the model for Fit 10 and 11.  The solid vertical lines show $r$ = 23$^{\prime\prime}$ and 166$^{\prime\prime}$, which is the same for both fits.  The blue dashed vertical lines show $r$ = 103$^{\prime\prime}$ and 126$^{\prime\prime}$ for Fit 10.  The green dash-dot vertical lines show $r$ = 95$^{\prime\prime}$, 109$^{\prime\prime}$, and 128$^{\prime\prime}$ for Fit 11.}
\end{centering}
\end{figure}

\section{DISCUSSION}

\subsection{Conclusions of the Test}

The results fail to provide evidence for the bar and spiral patterns having the same value of $\Omega_p$.  Most of the results show a value of $\Omega_p$ that is larger in the bar region than in the spiral pattern region.  Evidence for different values of $\Omega_p$ for bar and spiral patterns is also reported from applying the TW84 method by Rand \& Wallin (2004) for NGC 1068, Hernandez et al. (2005) for NGC 4321, and Chemin \& Hernandez (2009) for UGC 628.

The evidence for mode coupling of the bar and spiral patterns is unreliable and inconsistent.  The radial profile of $\Omega_p$ for Fit 2 shows a discontinuity at $r$ = 131$^{\prime\prime}$ where the CR of the larger value of $\Omega_p$ is approximately coincident with the $m$ = 2 ILR of the smaller value of $\Omega_p$.  This coincidence is unreliable because assumption (1) of the original TW84 method is violated.  The radial profile of $\Omega_p$ for Fit 10 shows a discontinuity at $r$ = 103$^{\prime\prime}$, near where the CR of the larger value of $\Omega_p$ is approximately coincident with the $m$ = 4 ILR of the smaller value of $\Omega_p$.  The coincidence is absent in Fit 11, which has the same value of BIC as Fit 10.   

The results are therefore the most consistent with the bar and spiral patterns being dynamically distinct features.  It is important to note that this conclusion is for the currently observed state of the H$\alpha$ emitting gas.  The H$\alpha$ bar and spiral patterns may have corotated, or been coupled by resonance, at some time in the past.  The nature of the spiral pattern is discussed further in Section 6.3.  

These conclusions could be interpreted as problematic for explaining why the spiral pattern appears to begin near the end of the bar. Sellwood \& Sparke (1988) show that this evidence for corotating bar and spiral patterns is unreliable. In their simulations they demonstrate that spiral patterns can appear to begin near the end of a faster rotating bar for most of the bar's rotation period, and that the offsets that do appear are small and infrequent. They also cite examples of galaxies where the beginning of the spiral pattern is clearly offset from the end of the bar.

\subsection{Properties of $\mathit \Omega_p$ for the Bar}

The results for Fit 10 and 11 are used for estimating the properties of the bar because they have the largest value of BIC, and are less biased than Fit 9.  They both show a constant value for $\Omega_p$ in the bar region that begins near an ILR and ends near a CR.  The values for the radius of CR, $R_{\mbox{\scriptsize CR}}$, are 94$^{\prime\prime}$ $\pm$ 2$^{\prime\prime}$ and 89$^{\prime\prime}$ $\pm$ 2$^{\prime\prime}$ for Fit 10 and Fit 11, respectively. The discontinuities in the models at $r$ = 103$^{\prime\prime}$ and 95$^{\prime\prime}$ for Fit 10 and 11, respectively, are close to the end of the constant trend in $\phi_{I2}$ for the K$_{\mbox{\footnotesize s}}$ band data.  

The radii of these discontinuities are an unlikely measure of the bar radius, $R_{\mbox{\footnotesize bar}}$, because they are inconsistent with the theoretical prediction that $R_{\mbox{\footnotesize bar}}$ $\leqslant$ $R_{\mbox{\scriptsize CR}}$ (e.g., Contopoulos 1980,  Tueben \& Sanders 1985).  A discontinuity in the model does not necessarily require that the true form of $\Omega_p$ has a discontinuity there.  For a continuous pattern, the method in Section 5.2.3 for locating discontinuities in the model will locate the radius of the largest gradient in $\Omega_p$.  The values of $R_{\mbox{\scriptsize CR}}$ for Fit 10 and 11 are therefore adopted as upper limits for $R_{\mbox{\footnotesize bar}}$.

\begin{deluxetable}{cccl}[]
\tablecaption{Comparison of Results for the Bar}
\startdata
\tableline\tableline
$R_{\mbox{\footnotesize bar}}$ & $R_{\mbox{\scriptsize CR}}$ & $\Omega_p$ & Reference \\
($^{\prime\prime}$) & ($^{\prime\prime}$) & (km s$^{-1}$ arcsec$^{-1}$) & \\
\tableline
\hskip -19pt  110 &  \hskip -22.5pt 100 & 3.1 $\pm$ 0.6 & T86\\
\nodata & \hskip 1pt 132 \hskip 1pt -- 145 & \nodata & OH89\\
\hskip -19pt  100 &  \hskip -22.5pt 138 & 2.1 $\pm$ 0.1 & JM95\\
\hskip 5pt 120 $\pm$ \hskip 3pt 10 &  \hskip -22.5pt 145 & \hskip -23.7pt 2.0 & L96\\
\hskip -19pt  100 & \hskip 1pt 113  $\pm$ \hskip 7pt 8 & 2.7 $\pm$ 0.2 & V01\\
\hskip 5pt 114  \hskip 1.5pt --   \hskip 0.5pt 127 &  \hskip -22.5pt 148 & \hskip -23.7pt 2.2 & Z08	\\
\tableline \tableline
\hskip 5pt 110 $\pm$ \hskip 3pt 10 & \hskip 1pt  130 \hskip 0pt $\pm$ \hskip 2.5pt 19 & 2.4 $\pm$ 0.5 & Mean \\
\tableline \tableline
$\leqslant$ 94 $\pm$ \hskip 7pt 2 & \hskip 5.0pt 94 $\pm$  \hskip 6.5pt 2 & 3.5 $\pm$ 0.1 & Fit 9	\\
 $\leqslant$ 89 $\pm$ \hskip 7pt 2 & \hskip 5.0pt 89 $\pm$  \hskip 6.5pt 2 & 3.7 $\pm$ 0.1 & Fit 10	
\enddata 
\end{deluxetable}

Table 5 summarizes previous estimates of $R_{\mbox{\footnotesize bar}}$, $R_{\mbox{\scriptsize CR}}$, and $\Omega_p$ for comparison with those in this paper.  The results for $\Omega_p$ are adjusted for the value of $\phi_i$ used in this paper, when different.  The previous estimates use different methods.  T86 assume the change in orientation of elliptical streamlines at $r$ = 15$^{\prime\prime}$ is an indication of ILR.  Ondrechen \& van der Hulst (1989, hereafter OH89) find that the change in the direction of the streaming velocities coincides with the dust lanes crossing the spiral arm.  This is consistent with a CR according to density wave theory, but OH89 note that this interpretation is inconsistent with $R_{\mbox{\footnotesize bar}}$ $\leqslant$ $R_{\mbox{\scriptsize CR}}$ if the bar and spiral patterns are corotating.  JM95 assumes the dip in the residual velocity field at $r$ = 173$^{\prime\prime}$ occurs at an $m$ = 4 OLR.  L96 and Z08 fit simulations of the gas to observations for a range of assumed values for $\Omega_p$.  V01 assumes a phase difference of zero for infrared and blue images is an indication of CR.

Most previous estimates for $R_{\mbox{\footnotesize bar}}$, $R_{\mbox{\scriptsize CR}}$, and $\Omega_p$ are inconsistent with those found in this paper.  Compared to the previous estimates, the values of $R_{\mbox{\footnotesize bar}}$ and $R_{\mbox{\scriptsize CR}}$ are smaller for Fit 10 and 11, and the values of $\Omega_p$ are larger.  The values found by T86 are the most similar to those found in this paper.  

The larger values of $\Omega_p$ that are measured in this paper for the bar are consistent with the residuals for the velocity field model found in Section 5.1. This can be shown using a model in the rotating frame of the galaxy for an elliptical orbit, 
\begin{eqnarray} \nonumber
\hskip -0pt V_e(r,\theta) = - V_{\theta,2}(r)\,\mbox{cos}(\theta)\,\mbox{cos}(2 \left[\theta - \phi_b \right]) \\ &\hskip -100 pt - V_{r,2}(r)\,\mbox{sin}(\theta)\,\mbox{sin}(2\left[ \theta - \phi_b \right]),
\end{eqnarray}
where that $V_{\theta,2}$ is the amplitude of the streaming in the azimuthal direction, $V_{r,2}$ is the amplitude of the streaming in the radial direction, and $\phi_b$ is the bar position angle found in Section 5.3 (Spekkens \& Sellwood 2007, Sellwood \& Sanchez 2010). The first term on the right of Equation (27) describes how material slows down in the azimuthal direction as it approaches the major axis of the ellipse, and speeds up as it approaches the minor axis of the ellipse. The second term on the right describes how material moves outwards in the radial direction as it approaches the major axis of the ellipse, and moves inward as it approaches the minor axis of the ellipse.

\begin{figure}[t!]
\begin{centering}
\includegraphics[width=0.395\textwidth]{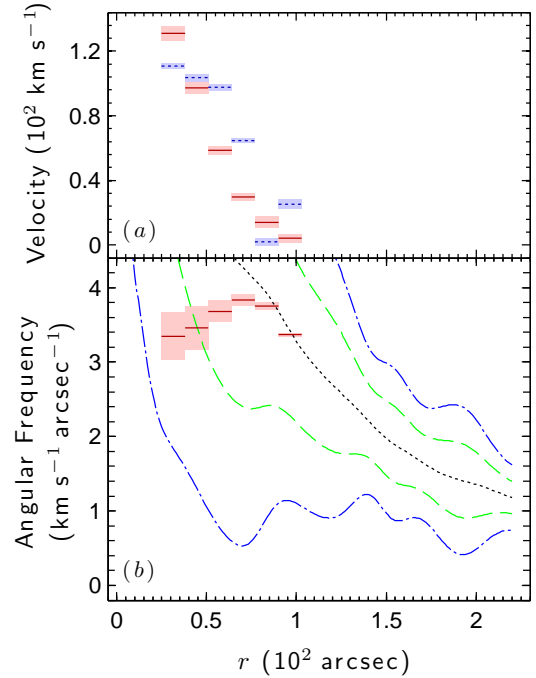} 
\caption{Estimate of $\Omega_p$ for the bar from the residuals for the velocity field model. In panel ($a$), the solid red line segments and light red shading show $V_{\theta,2}$ and 95\% CIs, respectively. Also in panel ($a$), the dotted blue line segments and light blue shading show $V_{r,2}$ and 95\% CIs, respectively. Panel ($b$) shows the calculated estimate of $\Omega_p$, and is formatted in the same way as in Figure 5.\\}
\end{centering}
\end{figure}

A crude estimate of $\Omega_p$ is found by fitting Equation (27) to the residuals for the velocity field model in the bar region, and calculating,
\begin{equation}
\Omega_p = \frac{V_\theta - V_{\theta,2}}{r}. 
\end{equation}
Fits of Equation (27) are found for the same rings used in the velocity field model that cover the bar. The results for 25$^{\prime\prime}$ $\leqslant$ $r$ $\leqslant$ 103$^{\prime\prime}$ are shown in Figure 13.  There are well defined trends in the values of $V_{\theta,2}$ and $V_{r,2}$. The mean of the six calculated $\Omega_p$ results is 3.7 $\pm$ 0.5 km s$^{-1}$ arcsec$^{-1}$. This is consistent with the values obtained for the bar in Fit 9 - Fit 11 using the TW84 method. 

\subsection{The Nature of the Spiral Pattern}

The radial profiles of $\Omega_p$ in the spiral pattern region most closely resemble what is expected for coupled spiral modes and tidal interactions.  The resemblance is only briefly discussed in this subsection because determining the nature of the spiral pattern is beyond the scope of this paper.  

The discontinuities at $r$ = 126$^{\prime\prime}$ and 128$^{\prime\prime}$ for Fit 10 and Fit 11, respectively, and 166$^{\prime\prime}$ for both fits, are consistent with mode coupling at $m$ = 4 ILRs.  It is unclear from the results, however, if the pattern is truly rigid at these resonances.  The assumption of a rigid pattern is a better established approximation for bars than it is for spiral patterns (e.g., Binney \& Tremaine 2008, Chapter 6).  

If the pattern is shearing, a discontinuity in the model may be where the gradient in the true form of $\Omega_p$ is the largest, as pointed out in Section 6.2.  In this case, the result for that region shows the mean value of $\Omega_p$.  This interpretation is consistent with the $\sim$1/$r$ behavior of $\Omega_p$ for the spiral pattern found by SW11.

If the spiral pattern is the result of a tidal interaction, the radial profile of $\Omega_p$ for the spiral pattern may resemble the results for Fit 9, or some combination, or average, of Fit 9 -- Fit 11.  There are no known companions to NGC 1365, but Roy \& Walsh (1997) find evidence for a recent merger, and Z08 find evidence for an infalling gas cloud.  These may have provided the tidal forcing that stimulated the spiral pattern, or helped maintain a spiral pattern that was stimulated by the bar at some time in the past.

\section{SUMMARY} 

Theories that attempt to explain the dynamical relationship between bar and spiral patterns in galactic disks make different predictions about the radial profile of the pattern speed.  These predictions are tested for the H$\alpha$ bar and spiral patterns of NGC 1365 by fitting mathematical models that are based on the TW84 method.  The findings are as follows.
\indent \begin{minipage}[ht!]{0.01\textwidth} \vskip -32pt 
(1)  
\end{minipage} 
\hskip 2pt \begin{minipage}[h!]{0.43\textwidth} \vskip 6.5pt
The results are inconsistent with a global wave mode or a manifold. The value of $\Omega_p$ decreases with increasing radius for the nuclear, bar, and spiral pattern regions of the galaxy. This violates the assumption of a single value for $\Omega_p$ in the original TW84 method.
\end{minipage}\\
\indent \begin{minipage}[ht!]{0.01\textwidth} \vskip -3pt 
(2)  
\end{minipage} 
\hskip 2pt \begin{minipage}[h!]{0.43\textwidth} \vskip 6.5pt
Violations of mass conservation in the continuity equation have a negligible effect on the results. 
\end{minipage}\\
\indent \begin{minipage}[ht!]{0.01\textwidth} \vskip -78pt 
(3)  
\end{minipage} 
\hskip 2pt \begin{minipage}[h!]{0.43\textwidth} \vskip 6.5pt
When $\Omega_p$ is allowed to vary with radius, the results for the general TW84 method are unstable to noise unless the ring width is at least a few times the resolution of the data, or a regularization greater than zeroth-order is used. First-order regularization provides the best defined corner in the L-curve criteria for selecting the amount of regularization. The results for first and second-order regularization are consistent for a pattern speed in the nuclear region that is larger than the value for the bar.
\end{minipage}\\
\indent \begin{minipage}[ht!]{0.01\textwidth} \vskip -12.5pt 
 {\color{white} (8)}
\end{minipage} 
\hskip 2pt \begin{minipage}[h!]{0.43\textwidth} \vskip 6.5pt
 {\color{white} The results for the spiral pattern are the most consistent with coupled spiral modes and tidal interactions.  }
\end{minipage}\\
\indent \begin{minipage}[ht!]{0.01\textwidth} \vskip -43pt 
(4)  
\end{minipage} 
\hskip 2pt \begin{minipage}[h!]{0.43\textwidth} \vskip 6.5pt
Excluding the nuclear region for a first-order regularized fit reveals a rigidly rotating bar. This demonstrates that the TW84 method can detect rigid patterns when allowed to vary with radius. A similar procedure fails to reveal a single value of $\Omega_p$ for a rigidly rotating spiral pattern.
\end{minipage}\\
\indent \begin{minipage}[ht!]{0.01\textwidth} \vskip -13pt 
(5)  
\end{minipage} 
\hskip 2pt \begin{minipage}[h!]{0.43\textwidth} \vskip 6.5pt
Selecting model discontinuities using BIC model selection also produces evidence for separate nuclear, bar, and spiral pattern speeds.
\end{minipage}\\
\indent \begin{minipage}[ht!]{0.01\textwidth} \vskip -52pt 
(6)  
\end{minipage} 
\hskip 2pt \begin{minipage}[h!]{0.43\textwidth} \vskip 6.5pt
The evidence for mode coupling of the bar and spiral patterns is unreliable and inconsistent. The results are the most consistent with the bar and spiral patterns being dynamically distinct features. This does not rule out the possibility that the bar and spiral patterns corotated, or were coupled by resonance, previous to the currently observed state of the H$\alpha$ emitting gas.
\end{minipage}\\
\indent \begin{minipage}[ht!]{0.01\textwidth} \vskip -43pt 
(7)  
\end{minipage} 
\hskip 2pt \begin{minipage}[h!]{0.43\textwidth} \vskip 6.5pt
The approximately constant pattern speed of the bar begins near the $m$ = 2 ILR and ends near the CR. The resulting values for $R_{\mbox{\scriptsize CR}}$ are smaller than most previous estimates, but consistent with the results for the potential minimum model.  The measured values of $\Omega_p$ for the bar are larger than most previous estimates.
\end{minipage}\\
\indent \begin{minipage}[ht!]{0.01\textwidth} \vskip -3pt 
(8)  
\end{minipage} 
\hskip 2pt \begin{minipage}[h!]{0.43\textwidth} \vskip 6.5pt
The results for the spiral pattern are the most consistent with coupled spiral modes and tidal interactions.  
\end{minipage}

\acknowledgements

The authors acknowledge the helpful comments of the referee  that  improved  this  paper.   The H$\alpha$ data are kindly provided by Ricardo Z\'anmar S\'anchez.  This research made use of the NASA/IPAC Extragalactic Database, which is operated by the Jet Propulsion Laboratory, California Institute of Technology, under contract with the National Aeronautics and Space Administration.

{}


\begin{thebibliography}{} \footnotesize

\bibitem[Akaike (1974)]{a74}Akaike H., 1974, ITAC, 19, 716
\bibitem[Aster et al. (2005)]{a05} Aster, R. C., Borchers, B., \& Thurber, C. H. 2005, Parameter Estimation and Inverse Problems (San Diego, CA: Elsevier Academic)
\bibitem[Athanassoula (2012)]{a12} Athanassoula, E. 2012, MNRAS, 426, 46
\bibitem[Bertin \& Arnouts (1996)]{ba96} Bertin, E., Arnouts, S. 1996, A\&AS, 117, 393
\bibitem[Bertin \& Lin (1996)]{bl96} Bertin, G., \& Lin, C. C. 1996, Spiral Structure in Galaxies: A Density Wave Theory (Cambridge, MA: MIT Press) 
\bibitem[Binney \& Tremaine (2008)]{bt08} Binney, J. \& Tremaine, S. 2008, Galactic Dynamics, 2nd ed. (Princeton, NJ: Princeton Univ. Press)
\bibitem[Buta et al. (2009)]{b09} Buta, R. J., Knapen, J. H., Elmegreen, B. G., et al. 2009, AJ, 137, 4487
\bibitem[Canzian (1998)]{c98} Canzian, B. 1998, ApJ, 502, 582
\bibitem[Chemin \& Hernandez (2009)]{ch09} Chemin, L., \& Hernandez, O. 2009, A\&A, 499, 25
\bibitem[Contopoulos (1980)]{c80} Contopoulos, G. 1980, A\&A, 81, 198
\bibitem[Debattista (2003)]{d03} Debattista, V. P. 2003, MNRAS, 342, 1194
\bibitem[Debattista \& Williams (2003)]{dw04} Debattista, V. P., \& Williams, T. B. 2004, ApJ, 605, 714
\bibitem[Dobbs \& Baba (2014)]{db14} Dobbs, C., \& Baba, J. 2014, PASA, 31, 35
\bibitem[Dobbs et al. (2010)]{d10} Dobbs, C. L., Theis, C., Pringle, J. E., \& Bate, M. R. 2010, MNRAS, 403, 625
\bibitem[Dubinski et al. (2008)]{d08} Dubinski, J., Gauther, J.-R., Widrow, L., \& Nickerson, S. 2008, in ASP Conf. Ser. 396, Formation and Evolution of Galaxy Disks, ed. J. G. Funes \& E. M. Corsini (San Francisco, CA: ASP), 321
\bibitem[Elmegreen \& Elmegreen (1982)]{ee82} Elmegreen, D. M., \& Elmegreen, B. G. 1982, MNRAS, 201, 1021
\bibitem[Emsellem et al. (2006)]{e06} Emsellem, E., Fathi, K., Herv\'e, W., et al. 2006, MNRAS, 365, 367
\bibitem[Emsellem et al. (2001)]{e01} Emsellem, E., Greusard, D., Combes, F., et al. 2001, A\&A, 368, 52
\bibitem[Engstr\"{o}m (1994)]{e94} Engstr\"{o}m, S. 1994, A\&A, 285, 801 (E94)
\bibitem[Fathi et al. (2007)]{f07} Fathi, K, Toonen, S., Faslc\'on-Barroso, J., et al. ApJ, 667, 137 
\bibitem[Fathi et al. (2009)]{f09} Fathi, K, Beckman, J. E., Hernandez, O., et al. ApJ, 704, 1657 
\bibitem[Feigelson \& Babu (2012)]{fb12} Feigelson, E. D., \& Babu, G. J. 2012, Modern Statistical Methods for Astronomy  (Cambridge, UK: Cambridge University Press)
\bibitem[Gabbasov et al. (2009)]{g09} Gabbasov, R. F., Repetto, P., \& Rosado, M. 2009, ApJ, 702, 392
\bibitem[Goldreich \& Lynden-Bell (1965)]{gl65} Goldreich, P., \& Lynden-Bell, D. 1965, MNRAS, 130, 125
\bibitem[Grand et al. (2012)]{g12} Grand, R. J. J., Kawata, D., \& Cropper, M. 2012, MNRAS, 426, 126
\bibitem[Hernandez et al. (2005)]{h05} Hernandez, O., Wozniak, H., Carignan, C., Amram, P., Chemin, L., \& Daigle, O. 2005, ApJ, 632, 253 
\bibitem[Huntley et al. (1978)]{h78} Huntley, J. M., Sanders, R. H., \& Roberts, W. W.. Jr. 1978, AJ, 221, 521
\bibitem[Jarrett et al. (2003)]{j03} Jarrett, T. H., Chester, T., Cutri. R., Schneider, S. E., \& Huchra, J. P. 2003, AJ, 125, 525

\bibitem[J\"{o}rs\"{a}ter \& van Moorsel (1995)]{JM95} J\"{o}rs\"{a}ter, S., \& van Moorsel, G. A. 1995, AJ, 110, 2037 (JM95) 
\bibitem[Julian \& Toomre (1966)]{jt66} Julian, W. H., \& Toomre, A. 1966, ApJ, 146, 810
\bibitem[Jungwiert et al. (1997)]{j97} Jungwiert, B., Combes, F., \& Axon, D. J. 1998, A\&AS, 125, 479
\bibitem[Kalnajs (1973)]{k73} Kalnajs, A. J. 1973, PASAu, 2, 174
\bibitem[Kass \& Raftery (1995)]{kr95} Kass, R. E., \& Raftery, A. E. 1995, JASA, 90, 773
\bibitem[Kawata et al. (2014)]{k14} Kawata, D., Hunt, J. A. S, Grand, R. J. J., Pasetto, S., \& Cropper, M. 2014, MNRAS, 443, 2757
\bibitem[Kormendy \& Norman (1979)]{kn79} Kormendy J., \& Norman C. A. 1979, ApJ, 233, 539
\bibitem[Kwok (2007)]{k07} Kwok, S. 2007, Physics and Chemistry of the Interstellar Medium (Sausailito, CA: University Science Books)
\bibitem[van der Kruit \& Allen (1978)]{vdka74} van der Kruit, P., \& Allen, R. 1978, ARA\&A, 16, 103
\bibitem[Kuchinski et al. (2000)]{k00} Kuchinski, L. E., Freedman, W. L., Madore, B. F., et al. 2000, ApJS, 131, 441
\bibitem[Laine et al. (2002)]{l02} Laine, S., Shlosman, I., Knapen, J. H., Peletier, R. F. 2002, ApJ, 567, 97
\bibitem[Lin \& Shu (1964)]{ls64} Lin, C. C., \& Shu, F. H. 1964, ApJ, 140, 646
\bibitem[Lin \& Shu (1966)]{ls66} Lin, C. C., \& Shu, F. H. 1966, PNAS, 55, 229
\bibitem[Lindblad (1963)]{l38} Lindblad, B. 1963, Stockholms Observatoriums Annaler, 22, 5
\bibitem[Lindblad (1956)]{l96} Lindblad, B. 1956, Stockholms Observatoriums Annaler, 19, 7
\bibitem[Lindblad et al. (1996)]{l96} Lindblad, P. A. B., Lindblad, P. O., \& Athanassoula, E. 1996, A\&A 313, 65 (L96)
\bibitem[Lindblad (1978)]{l78} Linblad, P. O. 1978, in Astronomical  Papers  Dedicated  to  Bengt Stro\'emgren, ed. A. Reiz \& T. Andersen (Copenhagen: Copenhagen University Observatory), 403
\bibitem[Lindblad (1999)]{l99} Lindblad, P. O. 1999, A\&ARv 9, 221
\bibitem[Meidt et al. (2009)]{m09} Meidt, S., Rand, \& R. J., Merrifield, M. 2009, ApJ, 702, 290
\bibitem[Meidt et al. (2008a)]{m08} Meidt, S., Rand, R. J., Merrifield, M., Debattista, V. P., \& Juntai, S. 2008a, ApJ, 676, 899
\bibitem[Meidt et al. (2008b)]{m08} Meidt, S., Rand, R. J., Merrifield, M., Shetty, R., \& Vogel, S. N. 2008b, ApJ, 688, 224
\bibitem[Merrifield et al. (2006)]{m06} Merrifield, M. R., Rand, R. J., \& Meidt, S. E. 2006, MNRAS, 366, 17 
\bibitem[Oh et al. (2008)]{so08} Oh, S. H., Kim, W. -T., \& Lee, H. M. 2008, ApJ, 683, 94
\bibitem[Ondrechen \& van der Hulst (1989)]{oh89} Ondrechen, M. P., \& van der Hulst, J. M. 1989, ApJ, 342, 29 (OH89)
\bibitem[P\'{i}nol-Ferrer et al. (2012)]{p12} P\'{i}nol-Ferrer, N., Lindblad, P. O., \& Fathi, K. 2012, MNRAS, 421, 1089
\bibitem[Ramsey \& Schafer (2002)]{rs02} Ramsey, F. L., \& Schafer, D. W. 2002, The Statistical Slueth (Pacific Grove, CA: Duxbury) 
\bibitem[Rand \& Wallin (2004)]{rw05} Rand, R. J., \& Wallin, J. F. 2004, ApJ, 614, 142
\bibitem[Roca-F{\`a}brega et al. (2013)]{r13}  Roca-F{\`a}brega, S., Valenzuela, O., Figueras, F. 2013, MNRAS, 432, 2878
\bibitem[Rohlfs (1977)]{r77} Rohlfs, K. 1977, Lectures on Density Wave Theory, (New York, NY: Springer-Verlag)
\bibitem[Romero-G\'omez et al. (2006)]{r06} Romero-G\'omez, M., Masdemont, J. J., Athanassoula, E., \& Garc\'ia-G\'omez, C. 2006, A\&A, 453, 39
\bibitem[Roy \& Walsh (1997)]{rw97} Roy, J.-R., \& Walsh, J. R. 1997, MNRAS, 288, 715
\bibitem[Salo et al. (2010)]{s10} Salo, H., Laurikainen, E., Knapen, J. H. 2010, ApJL, 715, 56
\bibitem[Sanders \& Huntley (1976)]{sh76} Sanders, R. H., \& Huntley, J. M. 1976, ApJ, 209, 53
555
\bibitem[Sandqvist et al. (1982)]{s82} Sandqvist A., J\"{o}rs\"{a}ter, S., \& Lindblad, P. O. 1982, A\&A 110, 336
\bibitem[Sellwood \& Carlberg (2014)]{sc14} Sellwood, J. A., \& Carlberg, R. G. 2014, ApJ, 785, 137
\bibitem[Sellwood \& Sparke (1988)]{ss88} Sellwood, J. A. \& Sparke, L. S. 1988, MNRAS, 231, 25P
\bibitem[Sellwood \& Wilkinson (1993)]{sw93} Sellwodd, J. A., \& Wilkinson, A. 1993, RPPh, 56, 173
\bibitem[Sellwood \& Zanmar Sanchez (2010)]{szs} Sellwood, J. A., \& Z\'anmar S\'anchez, R. 2010, MNRAS, 404, 1733
\bibitem[Schwarz (1978)]{s78} Schwarz, G. 1978, Ann. Stat, 6, 461
\bibitem[Schwarz (1981)]{s81} Schwarz, M. P. 1981, ApJ, 247, 77
\bibitem[Speights \& Westpfahl (2011)]{sw11} Speights, J. C., \& Westpfahl, D. J. 2011, ApJ, 736, 70 (SW11)
\bibitem[Speights \& Westpfahl (2012)]{sw12a} Speights, J. C., \& Westpfahl, D. J. 2012, ApJ, 752, 52
\bibitem[Spekkens \& Sellwood (2007)]{ss07} Spekkens, K., \& Sellwood, J. A. 2007, ApJ, 664, 204

\bibitem[Sygnet et al. (1988)]{s88} Sygnet, J. F., Tagger, M., Athanassoula, E., \& Pellat, R. 1988, MNRAS, 232, 733
\bibitem[Tagger et al. (1987)]{t87} Tagger, M., Sygnet, J. F., Athanassoula, E., \& Pellat, R. 1987, ApJL, 318, L43 
\bibitem[Toomre (1969)]{t69} Toomre, A. 1969, ApJ, 158, 899
\bibitem[Tremaine \& Weinberg (1984)]{tw84} Tremaine, S., \& Weinberg, M. D. 1984, ApJL, 282, L5 (TW84)
\bibitem[Tueben et al. (1986)]{t86} Tueben, P. J., Sanders, R. H., Atherton, P. D., \& van Alba, G. D. 1986, MNRAS 221, 1 (T86)
\bibitem[Tueben \& Sanders (1985)]{ts85} Teuben, P. J., \& Sanders, R. H. 1985, MNRAS, 212, 257
\bibitem[Tutukov \& Fedorova (2006)]{tf06} Tutukov, A. V., \& Fedorova, A. V. 2006, ARep, 50, 785
\bibitem[Vera-Villamizar et al. (2001)]{v01} Vera-Villamizar, N., Horacio, D., Puerari, I., \& de Carvalho, R. 2001, ApJ, 547, 187 (V01)
\bibitem[Wada et al. (2011)]{w11} Wada, K., Baba, J., \& Saitoh, T. R. 2011, ApJ, 735, 1
\bibitem[Westpfahl (1998)]{w98} Westpfahl, D. J. 1998, ApJS, 115, 203 
\bibitem[Z\'anmar S\'anchez et al. (2008)]{Z08} Z\'anmar S\'anchez , R., Sellwood, J. A., Weiner, B. J., \& Williams, T. B. 2008, ApJ, 674, 797 (Z08) 
\end{thebibliography}
\end{document}